\documentclass[%
 reprint,
superscriptaddress,
nofootinbib,
 amsmath,amssymb,
 aps,
 physrev,
longbibliography,
]{revtex4-2}

\usepackage{hyperref}
\usepackage{amsmath}
\usepackage{amssymb}
\usepackage{cleveref}
\usepackage{appendix}

\begin{document}

\title{Unified Evolution of Electromagnetic Sources in Homogeneous Fields}

\author{Ivanina Ilieva}
\email{ivanina.ilieva@kit.edu}
\affiliation{Institute of Theoretical Solid State Physics, Karlsruhe Institute of Technology, Kaiserstr. 12, 76131 Karlsruhe, Germany}
\author{Carsten Rockstuhl}
\affiliation{Institute of Theoretical Solid State Physics, Karlsruhe Institute of Technology, Kaiserstr. 12, 76131 Karlsruhe, Germany}
\affiliation{Institute of Nanotechnology, Karlsruhe Institute of Technology, Kaiserstr. 12, 76131 Karlsruhe, Germany}
\author{Ivan Fernandez-Corbaton}
\affiliation{Institute of Nanotechnology, Karlsruhe Institute of Technology, Kaiserstr. 12, 76131 Karlsruhe, Germany}

\date{July 23, 2025}

\begin{abstract}
In this work, we study the behavior of elementary electromagnetic sources, i.e., point-like electric charges and intrinsic magnetic dipoles, in the presence of homogeneous electromagnetic fields in a classical and covariant setting. We show that the respective evolution equations for both kinds of sources can be formulated using a single Lorentz-like transformation with suitably adjusted parameters, thereby unifying the fundamental behavior of these intrinsic particle properties. We arrive at this description by expressing the evolution of the electric sources, governed by the Lorentz force, as a series of infinitesimal boosts and rotations acting on them, with the electric and magnetic field as the corresponding parameters. Upon suitable adjustments, we find a new effective Lorentz-like transformation, applicable to both electric and magnetic sources. We provide the results using both the tensorial and pseudovector representation of the magnetic sources. Finally, we obtain a non-relativistic limit of the evolution equation for magnetic sources.
\end{abstract}


\maketitle
\section{Introduction and Summary}

Maxwell's equations tell us that electromagnetic fields are generated by certain objects such as electric charges and currents, the former of which we call elementary electric sources of Maxwell fields. However, these are not the only set of sources that cause and interact with electromagnetic fields. One may instead consider both electric charge and magnetic spin as the fundamental sources of Maxwell fields. The latter is connected to the intrinsic magnetic dipole moments of certain particles, which we call the magnetic sources. The behavior of the electric sources in homogeneous fields is described by the Lorentz force, which has a well-known covariant form. We give an overview of the classical description of magnetic sources in the following.

After Uhlenbeck and Goudsmit \cite{uhlenbeck_goudsmit} proposed the concept of an intrinsic rotation of electrons, called spin, in 1925, many authors contributed to the development of the theory of classical particles with spin. In 1926, Thomas \cite{Thomas1926} built on the idea and corrected it by accounting for relativistic effects on electrons moving in electromagnetic fields, thereby lifting the discrepancies in the initial spinning electron model. With the knowledge of Thomas' work, in that same year, Frenkel \cite{frenkel1926electrodynamics} published his theory on covariant spinning electron dynamics using a tensor to describe the intrinsic magnetic dipole moment. A few years later came Tamm \cite{tamm_1929}, who described spin covariantly via a pseudovector. In 1959, Bargmann, Michel, and Telegdi \cite{bmt_paper} published their famous paper on the covariant description of spin precession in homogeneous electromagnetic fields, an equation, which came to be known as the (T)BMT equation. A comprehensive review of the history of electron spin is presented in \cite{history_of_spin}. For more studies on covariant dynamics of particles with an intrinsic magnetic moment, see \cite{spin_bhabha_corben_1941,spin_corben_1961,VANHOLTEN19913,Bagrov_Classical_ST,Bordovitsyn_2001_SPINORBITAL,KUDRYASHOVA_natalia} and references therein. Particles with both intrinsic electric and magnetic dipole moment have also been studied, for example, in \cite{silenko_gamma_paper,fukuyama_silenko_tmbt_edm}. Relativistic spinning particle dynamics have also been explored using a Lagrangian and a Hamiltonian formalism for external electromagnetic fields \cite{Skagerstam_1981,Hoeneisen_ClassicalED,CHAICHIAN1997188}. The coupling of spin to gravitational fields in the framework of general relativity has been studied as well \cite{skagerstam_gravitationalfield,karlyee_spinning_gravitational,Pomeranskii:2000,mashhoon_spin_precession,ramirez,mashhoon_spinning_twistedGW, mashhoon_gravitomagneticSG,obukhov_spin_EM_G}. The advantage of classical spin models is that they tend to be more intuitive, while still providing an accurate description of spin properties.

For particles with spin in very strong or inhomogeneous fields additional terms arise, which originate from a Stern-Gerlach-type force. A covariant formulation of the Stern-Gerlach force was first considered by Frenkel \cite{frenkel1926electrodynamics} and then by Good \cite{good} and Nyborg \cite{nyborg1964approximate}, among others. It has been revisited and extended in recent years \cite{rafelski_covSG}, and the dynamics of both neutral and charged particles were studied in \cite{Formanek_covSG_neutral_extPW,Formanek_covSG_neutral_extlinearPW,formanek_covariantSG_charged_extPW}. We do not consider such contributions here since we focus on homogeneous fields and fields, whose gradients are negligibly small. Further extensions to the dynamical equations and implications thereof are presented in \cite{extended_bmt}. In the framework of general relativity, spin can also couple to gravitational fields and a gravitomagnetic Stern-Gerlach force arises (see, for example, \cite{mashhoon_gravitomagneticSG}).

In the present work, we are inspired by the similarities between the covariant Lorentz force, which applies to electric sources in homogeneous electromagnetic fields, and a Lorentz transformation acting on such electric sources. A close inspection of the explicit form of the Lorentz force shows two terms -- one of them describes how an electric field moves a charge along its field lines, and the other one portrays how the magnetic field rotates already moving charges around itself. We show that we can express the evolution of electric sources as a series of infinitesimal Lorentz transformations, which contain a boost and a rotation with the electric and magnetic fields as parameters, respectively. We hence adopt the perspective that the electromagnetic fields drive the electric sources to evolve by means of a Lorentz-like transformation. After carefully considering the application of such a transformation on the magnetic sources and a comparison with known evolution equations, we arrive at a modified operation. What we call an effective Lorentz-like transformation is not only applicable to the magnetic sources but also, in that exact form, to the electric ones. In both cases, the transformation yields the correct evolution equations for both kinds of sources, uniting their principle driving mechanisms in terms of a Lorentz-like transformation with effective electromagnetic fields as parameters.

This article is structured as follows. In Sec.~\ref{sec:notation_definitions}, we give a short introduction to the covariant notation for the sake of self-containment, and we provide the definitions of the relevant quantities and relations. Then, in Sec.~\ref{sec:el_sources}, we study the correspondence between the Lorentz force and a series of Lorentz-like transformations using the electromagnetic fields as parameters. In Sec.~\ref{sec:magn_sources} we apply such a Lorentz-like transformation on the magnetic sources, characterized by a second-rank tensor, and explore their evolution, which prompts us to reformulate the transformation in order for it to yield the correct evolution equations for both kinds of sources. The explicit form and non-relativistic approximations to different orders are discussed in Sec.~\ref{subsec:explicit_nonrel_magnetic_evolution}. Finally, we connect our description of the magnetic sources to one using a pseudovector in Sec.~\ref{subsec:spin_pseudovector_rep} and give the relations between the different quantities. Section~\ref{sec:conclusions} contains the conclusions that we draw from this work as well as an outlook.

\section{Notation and Definitions}
\label{sec:notation_definitions}
We work in the formalism of special relativity, which uses covariant notation. The central point is the mixing of space and time, which form a four-dimensional hyperbolic space, called Minkowski space-time with the metric tensor $\eta^{\mu\nu}=\eta_{\mu\nu}=diag(1,-1,-1,-1)$. We distinguish between covariant and contravariant elements, which are related to each other by the metric tensor. For example, the covariant and contravariant four-position vectors are given as
\begin{equation}
        x^\mu = \begin{pmatrix}
            ct \\ x \\ y \\ z
        \end{pmatrix} = \begin{pmatrix}
            ct\\ \mathbf{r}
        \end{pmatrix}\ \ \ \ \text{and} \ \ \ \  x_\mu = \eta_{\mu\nu}x^\nu = \begin{pmatrix}
        ct \\ -\mathbf{r}
    \end{pmatrix} \ ,
\end{equation}
respectively, with $c$ being the speed of light. The Greek indices run over $\{ 0,1,2,3 \}$, and the Einstein summation convention is used for contraction between a co- and contravariant index. The invariant distance element in relativity is the proper time $\tau$, defined by 
\begin{equation}
\label{invariant_volume_element}
    c^2 \mathrm{d}\tau ^2 = \eta_{\mu\nu}\mathrm{d}x^\mu \mathrm{d}x^\nu = c^2 \mathrm{d}t^2 - \mathrm{d}x^2 - \mathrm{d}y^2 - \mathrm{d}z^2 \ ,
\end{equation}
which corresponds to time as experienced in the rest frame of the object. The proper time is typically used to parametrize the trajectory of a point particle, and hence, all variables connected to its movement are implicit functions of $\tau$. In this work, we frequently refer to an object's four-velocity
\begin{equation}
    u^\mu = \frac{\mathrm{d}{x}^\mu}{\mathrm{d}\tau} = \begin{pmatrix}
        \gamma c \\ \gamma \mathbf{v}
    \end{pmatrix} \, ,
\end{equation}
where $\mathbf{v}$ is the three-velocity vector of the particle, and the relativistic factor appearing is defined as $\gamma^{-1}=\sqrt{1-\beta^2}$ with $\pmb{\beta}=\frac{\mathbf{v}}{c}$. A Lorentz scalar is a constant in any reference frame, and in the case of the projection of the four-velocity onto itself, that value is $u_\mu u^\mu = c^2$. Taking the derivative with respect to proper time on both sides yields the condition
\begin{equation}
\label{eq:u_udot=0}
    u_\mu \frac{\mathrm{d}{u}^\mu}{\mathrm{d}\tau} = 0 \,,
\end{equation}
that is, the four-velocity $u^\mu$ and four-acceleration $\frac{\mathrm{d}{u}^\mu}{\mathrm{d}\tau}$ are orthogonal in the four-dimensional hyperbolic Minkowski space. 

The next important object of discussion is the totally antisymmetric second-rank tensor consisting of the electric and magnetic fields, i.e., the electromagnetic tensor
\begin{equation}
    F^{\mu\nu} = \begin{pmatrix}
        0 & -E_x /c & -E_y /c & - E_z /c \\
         E_x /c & 0 & -B_z & B_y \\
        E_y /c & B_z & 0 & -B_x \\
         E_z /c & -B_y & B_x & 0 
    \end{pmatrix} \, ,
\end{equation}
which is present in the covariant form of Maxwell's equations. Based on how the polarization and magnetization density, $\mathbf{P}$ and $\mathbf{M}$, respectively, enter Maxwell's equations, we can see that they also build a tensor with $\mathbf{E}/c \rightarrow -c\mathbf{P}$ and $\mathbf{B}\rightarrow \mathbf{M}$ \cite[Chapter 11.10]{jackson1999classical}. However, here we will focus on point particles and their intrinsic magnetic moments. One such moment $\pmb{\mu}$ gives rise to an electric dipole moment $\pmb{d}$ when considered from an inertial reference frame, moving with a constant velocity with respect to the reference frame, in which the particle hosting $\pmb{\mu}$ is at rest. We know that $\pmb{\mu}$ must be part of the space-space components of the moment tensor $\Sigma ^{\mu\nu}$ and $\pmb{d}$ of the space-time ones. Although typically the literature on the matter of the $\Sigma ^{\mu\nu}$ tensor dynamics (such as \cite{frenkel1926electrodynamics} and works based on it) consists of just the electric and magnetic dipole moments, we cannot {\it a priori} assume that they transform the same way as the densities $\mathbf{P}$ and $\mathbf{M}$ and that we can just substitute the densities by the moments. As is pointed out in \cite{silenko_gamma_paper}, one can determine the transformation properties of the moments by the following consideration. The magnetization density $\mathbf{M}$ is the magnetic moment per unit volume. The connection between the magnetic dipole moment $\pmb{\mu}$ and the magnetization density $\mathbf{M}$ reads as
\begin{equation}
\label{magnetization_density_vs_moment}
    \mathbf{M} = \frac{\mathrm{d}\pmb{\mu}}{\mathrm{d}V} \, .
\end{equation}
We already know that $\mathbf{M}$ transforms as the space-space part of a tensor under a Lorentz transformation, e.g., a boost. The volume element $\mathrm{d}V=\mathrm{d}x\mathrm{d}y\mathrm{d}z$ becomes $\mathrm{d}V'=\mathrm{d}x'\mathrm{d}y'\mathrm{d}z'$ in a boosted frame. Without loss of generality, we can align the spatial axes such that the boost is in the $x$-direction. Then, only the $x$-dimension is Lorentz contracted with $\mathrm{d}x'=\frac{1}{\gamma}\mathrm{d}x$ and $\mathrm{d}y'\mathrm{d}z' = \mathrm{d}y\mathrm{d}z$. Hence, we can write, in general, that
\begin{equation}
    \mathrm{d}V=\frac{1}{\gamma}\mathrm{d}V_{\text{rest}} \, .
\end{equation}
Upon comparison of the magnetization density in the rest frame and in the boosted frame, 
\begin{equation}
    \mathbf{M}_\text{rest} = \frac{\mathrm{d}\pmb{\mu}
    _\text{rest}}{ \mathrm{d}V_\text{rest}} \ , \ \ \mathbf{M} = \frac{\mathrm{d}\pmb{\mu}}{\mathrm{d}V} = \frac{ \gamma\mathrm{d}\pmb{\mu}}{\mathrm{d}V _\text{rest}} \ ,
\end{equation}
we see that, since the left-hand sides ($\mathbf{M}$) transform as the space-space part of an antisymmetric tensor and the denominators on the right-hand sides ($\mathrm{d}V _\text{rest}$) are identical, the numerators ($\gamma\mathrm{d}\pmb{\mu}$) also transform like the space-space part of an antisymmetric tensor. By analogous considerations, one finds that $c\gamma \mathrm{d}\pmb{d}$ is the space-time part of the same tensor. The tensor for the electric and magnetic dipole moment of a point particle is hence given by
\begin{equation}
\label{sigma_def}
    \Sigma ^{\mu\nu} = \gamma \begin{pmatrix}
        0 & c d_x & c d_y & c d_z \\
        - c d_x & 0 & -\mu _z & \mu _y \\
         -c d_y & \mu _z & 0 & -\mu _x \\
        -c d_z & -\mu _y & \mu _x & 0 
    \end{pmatrix} \ ,
\end{equation}
which we call the electromagnetic moment tensor. The additional $\gamma$ factor in the tensor is important because it is frame-dependent. Therefore, such a factor changes the transformation properties of $\pmb{\mu}$ and $\pmb{d}$ upon a Lorentz boost from the rest frame to a different reference frame moving with velocity $\mathbf{v} $ with respect to the rest frame, as follows:
\begin{equation}
\label{mu_d_transformation_props}
    \begin{aligned}
        \pmb{\mu} &= \pmb{\mu}_\text{rest} - \mathbf{v}\times \pmb{d}_\text{rest} - \frac{1}{c^2} \frac{\gamma}{\gamma +1}\mathbf{v}(\mathbf{v}\cdot\pmb{\mu}_\text{rest}) \ , \\
        \pmb{d} &= \pmb{d}_\text{rest} + \frac{1}{c^2} \mathbf{v}\times \pmb{\mu}_\text{rest} - \frac{1}{c^2} \frac{\gamma}{\gamma +1}\mathbf{v}(\mathbf{v}\cdot\pmb{d}_\text{rest}) \, .
    \end{aligned}
\end{equation}
Currently, there is no experimental evidence of the existence of an intrinsic electric dipole moment (EDM) in fundamental particles. For a comprehensive review of the subject, see \cite{edm_paper} and references therein. We will assume that the point particles that we consider do not have an EDM at rest and, in turn, that the electric moment $\pmb{d}$ appearing in the electromagnetic moment tensor $\Sigma^{\mu\nu}$ arises only as a consequence of evaluating it in a reference frame, which is moving with a given instantaneous velocity with respect to the particle rest frame. Hence, at rest, $\Sigma^{\mu\nu}$ consists only of space-space components corresponding to the intrinsic magnetic moment of the particle $\pmb{\mu}_\text{rest}$. With $\pmb{d}_\text{rest}=0$ in Eq.~(\ref{mu_d_transformation_props}), we get an additional condition relating the electric and magnetic moment of a particle. The connection reads as
\begin{equation}
\label{no_EDM_rest_explicit}
    \pmb{d} = \frac{1}{c^2} \mathbf{v}\times \pmb{\mu}_\text{rest} \ ,
\end{equation}
or equivalently,
\begin{equation}
    \label{no_EDM_rest} \Sigma^{\mu\nu}u_\nu = 0
\end{equation}
in covariant form, the latter of which is known as the Frenkel condition \cite{frenkel1926electrodynamics}. For more details on the equivalence of Eqs.~(\ref{no_EDM_rest_explicit}) and (\ref{no_EDM_rest}), see Appendix~\ref{appendixA_Eqs11_12}. We need such a constraint to eliminate the degrees of freedom of the tensor that one would otherwise associate with an electric moment independent of the magnetic one. The electromagnetic moment tensor $\Sigma^{\mu\nu}$ has two remaining degrees of freedom -- three out of the six initial ones are taken care of by way of Eq.~(\ref{no_EDM_rest}) and one is removed by the fixed value of $\mu = |\pmb{\mu}_\text{rest}|$ via the Lorentz invariant $\Sigma^{\mu\nu}\Sigma_{\mu\nu}=-2\mu$.

\section{Evolution of the Electric Sources}
\label{sec:el_sources}
We start by analyzing what drives the elementary electric sources and how. We are interested in the trajectory of an electrically charged point particle, i.e., an electric monopole, in the presence of homogeneous electromagnetic fields and fields with negligibly small gradients. The four-velocity is the tangent vector to the particle trajectory at a certain proper time. Hence, given the necessary initial conditions, the change of the four-velocity of a particle determines its trajectory. The evolution of the four-velocity of an electric charge can, therefore, be thought of as the evolution of the charge itself. It is given by the force acting on it in an external electromagnetic field -- the well-known Lorentz force, here in covariant form
\begin{equation} \label{lorentz_force_covariant}
    m\frac{\mathrm{d}{u}^\mu}{\mathrm{d}\tau} = e F^\mu _{\ \nu} u^\nu \, ,
\end{equation}
where $m$ is the mass and $e$ the elementary charge of the particle. If field gradients are non-negligible, a coupling of the particle trajectory to the magnetic moment arises, i.e., the latter influences the former. Such contributions are given in \cite[Eq.~(6)]{Formanek_covSG_neutral_extPW}, but we do not consider them here. It is worth noting that in the setting of homogeneous fields, we can treat the field tensor as just another constant and solve the ordinary differential equation for $u^\mu$ given by the Lorentz force, Eq.~(\ref{lorentz_force_covariant}). The expression for the trajectory then contains an exponentiation of the proper time and the electromagnetic field tensor (see \cite[Chapter~1.1.3]{Itzykson_Zuber_QFT}). To gain more insight, we rewrite Eq.~(\ref{lorentz_force_covariant}) explicitly in its temporal and spatial components
\begin{equation} \label{lorentz_force_noncovariant}
    \frac{\mathrm{d}}{\mathrm{d}\tau}\begin{pmatrix}
        \gamma c \\ \gamma \mathbf{v}
    \end{pmatrix} = \frac{e}{m} \begin{pmatrix} \gamma
        \mathbf{E}\cdot\mathbf{v} \\ \gamma \mathbf{E} + \gamma \mathbf{v} \times  \mathbf{B}
    \end{pmatrix}  \, .
\end{equation}
The right-hand side of the above equation can be interpreted as follows. The spatial components give the three-dimensional Lorentz force acting on the charge -- the \textit{direction} in which and the \textit{strength} with which it is being pushed and pulled by the fields -- telling us how its velocity changes and thereby determining its trajectory. The time component represents the work exerted on the change by the electric field, i.e., the rate of change of the particle's energy \cite[Chapter 11.9]{jackson1999classical}. The $\gamma$-factor coming from the four-velocity is present in all components and ensures that the resulting four-vector transforms correctly under the transformations of special relativity. 

We want to analyze Eq.~(\ref{lorentz_force_noncovariant}) with some more care. The spatial part describes how the velocity associated with the motion of an electrically charged point particle changes ($\frac{\mathrm{d}}{\mathrm{d}\tau} \gamma \mathbf{v}$). An electric field makes such a charge {\it move along} it ($\propto\gamma\mathbf{E}$), while a magnetic field causes an already moving charge to {\it rotate around} it ($\propto\gamma\mathbf{v}\times\mathbf{B}$). How the temporal component, i.e., the energy of the charge, changes is dictated by the part of the three-velocity parallel to the electric field. With these considerations, a similarity emerges between the Lorentz force and a Lorentz boost, which \textit{moves} a system \textit{along} a given direction with a certain rapidity, combined with a rotation, which \textit{rotates} the system \textit{around} an axis by some angle. By drawing this parallel, we can think of the electromagnetic field as driving the sources to evolve by way of a Lorentz-like transformation. 

A system experiencing acceleration, such as the one where a Lorentz force acts on the electric sources, is non-inertial. Lorentz boosts only relate coordinate systems moving with a constant velocity relative to each other, which is not the case for the above consideration. However, we may still describe an accelerated motion as a series of infinitesimal Lorentz boosts with a constantly adjusted rapidity parameter. We shall examine the idea of the electromagnetic field driving the sources through a Lorentz-like transformation containing a boost and a rotation by applying such an infinitesimal boost and rotation to $u^\mu$ and comparing the results. In the infinitesimal regime, the two transformations commute in their general forms up to the first order. The active transformation for an infinitesimal Lorentz boost combined with an infinitesimal rotation is given by \cite[Chapter 11.7]{jackson1999classical}
\begin{equation}
\label{active_infin_Lorentz_transformation}
\begin{aligned}
    L ^\mu  _{\ \nu} & = \delta ^\mu _\nu + (\pmb{\varphi}\cdot \mathbf{J} + \pmb{\zeta}\cdot\mathbf{K})^\mu  _{\ \nu} \\
    &= \delta ^\mu _\nu + \begin{pmatrix}
        0 & \zeta _x & \zeta _y & \zeta _z \\
        \zeta _x & 0 & -\varphi _z & \varphi _y \\
        \zeta _y & \varphi _z & 0 & -\varphi _x \\
        \zeta _z & -\varphi _y & \varphi _x & 0 \\
    \end{pmatrix}
\end{aligned}
\end{equation}
with the typical transformation parameters -- the three angles of rotation represented by $\pmb{\varphi}$ and the rapidity vector $\pmb{\zeta}= \hat{\mathbf{v}}_b\tanh ^{-1}\frac{v_b}{c}$ of the boost with $\hat{\mathbf{v}}_b=\frac{\mathbf{v}_b}{v_b}$ and $v_b=\sqrt{\mathbf{v}_b\cdot\mathbf{v}_b}$ -- and their corresponding generators -- the vectors of matrices $\mathbf{J}$ and $\mathbf{K}$, respectively, the explicit form of which is given in Appendix~\ref{appendixB1_infLT}. Applying this Lorentz transformation to the four-velocity $u^\mu$ yields the transformed quantity in covariant form and explicitly for the components
\begin{equation}
\label{u'}
\begin{aligned}
    {u' }^\mu &= L ^\mu  _{\ \nu} u^\nu , \\
    \begin{pmatrix}
        \gamma' c \\ (\gamma \mathbf{v})'
    \end{pmatrix} &= \begin{pmatrix}
        \gamma c \\ \gamma\mathbf{v}
    \end{pmatrix} + \begin{pmatrix}
        \pmb{\zeta}\cdot (\gamma \mathbf{v}) \\ \gamma c \pmb{\zeta} + \pmb{\varphi}\times (\gamma\mathbf{v})
    \end{pmatrix} \ .
\end{aligned}
\end{equation}
To compare the above equation to the Lorentz force for a four-velocity, Eq.~(\ref{lorentz_force_covariant}), we modify the latter first. Using the limit representation of a derivative, here with respect to proper time, $\frac{\mathrm{d}}{\mathrm{d}\tau} f(\tau) = \lim _{\Delta \tau \rightarrow 0} \frac{f (\tau + \Delta \tau) - f (\tau) }{\Delta \tau}$, we get in the limit $\Delta\tau\rightarrow 0$
\begin{equation}\label{u_delta_tau}
\begin{aligned}
    u^\mu (\tau + \Delta\tau) &= u^\mu (\tau) + \Delta\tau\frac{e}{m} F^\mu  _{\ \nu} u^\nu (\tau) \\
    &= \Big( \delta ^\mu _\nu + \Delta\tau\frac{e}{m} F^\mu _{\ \nu} \Big) u^\nu (\tau) \, , \\
    \begin{pmatrix}
        \gamma (\tau + \Delta\tau)c \\ (\gamma \mathbf{v} )(\tau + \Delta\tau)
    \end{pmatrix} &= \begin{pmatrix}
        \gamma (\tau)c \\ (\gamma \mathbf{v} )(\tau)
    \end{pmatrix} \\ & \ \ \ + \Delta\tau \frac{e}{m} \begin{pmatrix} \frac{1}{c}
        \mathbf{E}\cdot(\gamma\mathbf{v}) \\ \gamma c \frac{1}{c}\mathbf{E} - \mathbf{B} \times  (\gamma\mathbf{v})
    \end{pmatrix}  \ 
\end{aligned}
\end{equation}
for the covariant and explicit forms. The above equation describes how the four-velocity after an infinitesimal increment in $\tau$ depends on that at the current $\tau$. Specifically, the second line in Eq.~(\ref{u_delta_tau}) is already reminiscent of an infinitesimal transformation acting on the four-velocity. It consists of an identity operation, $\delta^\mu _\nu$, and an additional term proportional to a small parameter, $\Delta \tau$, and a matrix operator, $F^\mu  _{\ \nu}$. 

We are now almost ready to compare Eqs.~(\ref{u'}) and (\ref{u_delta_tau}). To bring it all together, we make the following argument: $u^\mu (\tau + \Delta\tau)=\big( \gamma (\tau + \Delta\tau)c ,\, (\gamma\mathbf{v}) (\tau + \Delta\tau) \big)$ in Eq.~(\ref{u_delta_tau}) represents how the four-velocity has changed after an infinitesimal proper time interval $\Delta\tau$, while ${u' }^\mu =\big( \gamma ' c , \, (\gamma\mathbf{v})' \big)$ in Eq.~(\ref{u'}) is the four-velocity after an infinitesimal Lorentz transformation consisting of a boost with rapidity $\pmb{\zeta}$ and a rotation by the angles in $\pmb{\varphi}$. With the consideration that the electromagnetic field drives the electric sources, i.e., their four-velocity, by way of boosting and rotating it, we take 
\begin{equation}
\label{identification_Deltau_isu'}
    {u' }^\mu \rightarrow u^\mu (\tau + \Delta\tau)   \ .
\end{equation}
We can then compare the respective right-hand sides of the explicit forms in Eqs.~(\ref{u'}) and (\ref{u_delta_tau}). The first term -- the four-velocity -- is the same in both equations. In the respective second terms, very similar expressions are found; in the temporal components of both, there is a scalar product with $\gamma\mathbf{v}$ and in the spatial ones -- a term proportional to $\gamma c$ and one consisting of a cross product with $\gamma\mathbf{v}$. We can now directly identify the remaining quantities and make the following substitution of the Lorentz transformation parameters
\begin{equation}
\label{parameter_substitution}
        \pmb{\zeta} \rightarrow  \Delta\tau\frac{e}{m} \frac{\mathbf{E}}{c} \ , \
        \pmb{\varphi} \rightarrow -\Delta\tau\frac{e}{m} \mathbf{B} \ .
\end{equation}
The identification is in agreement with our previous discussion -- the electric field boosts the sources along itself and corresponds to the boost parameter $\pmb{\zeta}$, while the magnetic field causes a rotation and corresponds to the rotation angles $\pmb{\varphi}$. We can now substitute the new parameters in the Lorentz transformation in Eq.~(\ref{active_infin_Lorentz_transformation}) to get the new Lorentz-like transformation in terms of the electromagnetic fields, which we call $\Lambda ^\mu _{\ \nu} $. It reads as
\begin{equation}
\label{new_transformation}
    \begin{aligned}
    \Lambda ^\mu _{\ \nu} & = \delta ^\mu _\nu + \Delta\tau\frac{e}{m} \begin{pmatrix}
        0 & E_x /c & E_y /c & E_z /c \\
         E_x /c & 0 & B_z & -B_y \\
        E_y /c & -B_z & 0 & B_x \\
         E_z /c & B_y & -B_x & 0 
    \end{pmatrix}  \\
    & = \delta ^\mu _\nu + \Delta\tau\frac{e}{m} F^\mu  _{\ \nu} \ .
\end{aligned}
\end{equation}
We observe that the parameter substitution reconstructs exactly the electromagnetic tensor. The above matrix is our Lorentz-like transformation driving the electric sources. The inverse is given by
\begin{equation}
\label{inverse_of_new_transformation}
    \Lambda _{\nu } ^{\ \mu} = (\Lambda ^{-1}) ^\mu  _{\ \nu} = \delta ^\mu _\nu - \Delta\tau\frac{e}{m} F^\mu  _{\ \nu} \ .
\end{equation}
Like any Lorentz transformation, this one keeps the metric tensor invariant, which is essential for the Lorentz invariance of the infinitesimal four-volume element related to the proper time in Eq.~(\ref{invariant_volume_element}). More details regarding the form of the inverse and the proof of the above statements can be found in Appendices~\ref{appendixB2_inversetranformation} and~\ref{appendixB3_metrictensorinvariance}.

\section{Evolution of the Magnetic Sources}
\label{sec:magn_sources}
After analyzing the Lorentz force that drives the electric sources to evolve in the presence of homogeneous electromagnetic fields, we aim to study the behavior of the magnetic sources under such conditions in the following. Although we discuss the electric and magnetic sources in a manner that might suggest that they are separate objects, it is important to note that this is often not the case. For example, all electrically charged fermions, such as the electron, possess an intrinsic half-integer spin, making them both electric and magnetic sources simultaneously. However, since we only consider the dynamics of the sources in homogeneous fields, we will treat it separately and will not assume any {\it a priori} coupling between the evolution of the electric sources (the trajectory of the particle) and the magnetic ones (the orientation of the magnetic moment)\footnote{In the case of fields, whose gradients are non-negligible, a coupling between the electric and magnetic source dynamics arises (see \cite[Eq.~(6)]{Formanek_covSG_neutral_extPW}) and we would not be allowed to assume independence here.}. The only requirement we impose is Eq.~(\ref{no_EDM_rest_explicit}), or equivalently, Eq.~(\ref{no_EDM_rest}), which ensures that there is no EDM in the instantaneous rest frame of the particle ($\mathbf{v}_\text{rest}=0$).

The magnetic sources are described using the totally antisymmetric second-rank tensor given in Eq.~(\ref{sigma_def}) together with the Frenkel condition, Eq.~(\ref{no_EDM_rest}). We know that while electric sources at rest are only affected by the electric field around them according to the Lorentz force, Eq.~(\ref{lorentz_force_noncovariant}), magnetic moments at rest precess around the magnetic field and do not couple to the electric one. There may not be an EDM at any point in the instantaneously accompanying rest frame. We will use these ground truths of the rest frame dynamics to check the correctness of the results we get. 

In Sec.~\ref{sec:el_sources}, we formulated the evolution of the electric sources as a series of infinitesimal boosts and rotations using the electric and magnetic fields as parameters, respectively. We shall assume {\it a priori} that the magnetic sources also evolve similarly in the presence of electromagnetic fields. To that end, we want to apply the same type of infinitesimal Lorentz-like transformation on the magnetic moment tensor $\Sigma^{\mu\nu}$. Since we are handling the electric and magnetic sources separately, we replace the constant $e/m$ in the infinitesimal term proportional to $\Delta\tau$ by a generalized one, $\kappa$, allowing its value to be different, while its units remain unchanged. We apply the Lorentz-like transformation to the magnetic moment tensor following the transformation rules for second-rank tensors to get
\begin{equation} \label{sigma'_transf}
\begin{aligned}
    {\Sigma'} ^{\mu\nu} &= \Lambda ^\mu _{\ \alpha}
    \Lambda ^\nu _{\ \beta}  \Sigma^{\alpha\beta}   \\
    &= \Sigma^{\mu\nu} + \Delta\tau\kappa \big( F^\mu  _{\ \alpha}\Sigma^{\alpha\nu} - F^\nu  _{\ \alpha}\Sigma^{\alpha\mu} \big) \ 
\end{aligned}
\end{equation}
up to first order in $\Delta\tau$. For more details on the above calculation, see Appendix~\ref{appendixC1_LlikeTonMomentTensor}. Reconstructing the proper time derivative using the same assumption as for the electric sources, ${\Sigma'} ^{\mu\nu} \rightarrow {\Sigma} ^{\mu\nu} (\tau + \Delta \tau)$ and accordingly ${\Sigma} ^{\mu\nu} = {\Sigma} ^{\mu\nu} (\tau)$, in the above equation yields
\begin{equation} \label{eq:sigma_dot_commutator}
\begin{aligned}
    \frac{\mathrm{d}\Sigma^{\mu\nu} }{\mathrm{d}\tau} &= \kappa \big( F^\mu  _{\ \alpha}\Sigma^{\alpha\nu} - F^\nu  _{\ \alpha}\Sigma^{\alpha\mu} \big) \\
    &= \kappa F^{[ \mu} _{\ \ \alpha} \Sigma ^{\alpha\nu ]} \\
    &= \kappa [F,\Sigma]^{\mu\nu} \ .
\end{aligned}
\end{equation}
Importantly, the acquired evolution equation yields a totally antisymmetric tensor on both hand sides. It is written in three different but equivalent forms that we can compare to the evolution equations for classical intrinsic spin found in literature, which were obtained by different means. The expressions in the three lines in Eq.~(\ref{eq:sigma_dot_commutator}) correspond to those in \cite{Bagrov_Classical_ST}, \cite{Bordovitsyn_2001_SPINORBITAL} and \cite{VANHOLTEN19913}, respectively. Comparison with the evolution equations in \cite{Bagrov_Classical_ST,Bordovitsyn_2001_SPINORBITAL, Hoeneisen_ClassicalED} yields the explicit expression for the prefactor $\kappa$ \footnote{The given $\kappa$ has been adjusted from the one in Refs. \cite{Bagrov_Classical_ST,Bordovitsyn_2001_SPINORBITAL}, which have an additional factor of $c$ in the denominator due to a difference in the definitions of the electromagnetic field and moment tensor elements.}
\begin{equation}
    \kappa = \frac{eg}{2m}\ ,
\end{equation}
where $g$ is the gyromagnetic ratio of the particle under consideration. Note that if we had a particle with $g=2$, the above constant would match that in the Lorentz-like transformation of the electric sources exactly ($\kappa=e/m$).

Now that we have arrived at an evolution equation for the electric and magnetic moments, Eq.~(\ref{eq:sigma_dot_commutator}), we need to check that it satisfies the condition that magnetic moments at rest only couple to magnetic fields and that it is in agreement with the Frenkel condition,  $\Sigma^{\mu\nu}u_\nu = 0$ in Eq.~(\ref{no_EDM_rest}). The explicit form of Eq.~(\ref{eq:sigma_dot_commutator}) reads as
\begin{equation}
    \begin{aligned}
        \frac{\mathrm{d} (\gamma \pmb{\mu})}{\mathrm{d}\tau} &= \frac{eg}{2m} \gamma \pmb{d}\times\mathbf{E} + \frac{eg}{2m} \gamma\pmb{\mu}\times\mathbf{B} \ , \\
        \frac{\mathrm{d} (\gamma\pmb{d})}{\mathrm{d}\tau} &= \frac{eg}{2m}\gamma \pmb{d}\times\mathbf{B} - \frac{eg}{2mc^2}\gamma\pmb{\mu}\times \mathbf{E} \ .
    \end{aligned}
\end{equation}
In the instantaneously accompanying rest frame with $\pmb{d}_\text{rest}=0$ (as per Eq.~(\ref{no_EDM_rest_explicit}) with $\mathbf{v}_\text{rest}=0$) and $\gamma_\text{rest}=1$ the remaining terms are
\begin{equation}
\label{ddot_mudot_rest_incorrect}
    \begin{aligned}
        \bigg( \frac{\mathrm{d}\pmb{\mu}}{\mathrm{d}\tau} \bigg) \bigg| _\text{rest} &= \frac{eg}{2m} (\pmb{\mu}\times\mathbf{B})|_\text{rest} \ , \\
        \bigg( \frac{\mathrm{d}\pmb{d}}{\mathrm{d}\tau}  \bigg) \bigg| _{\text{rest}}&= - \frac{eg}{2mc^2}(\pmb{\mu}\times \mathbf{E})|_\text{rest} \ .
    \end{aligned}
\end{equation}
The evolution equation of the magnetic moment in the rest frame does fulfill the condition that it may only couple to the magnetic field. The change of the electric moment is not identically zero at rest for a general electric field, which authors find troubling \cite{frenkel1926electrodynamics,Bagrov_Classical_ST}. To check if the non-vanishing derivative with respect to $\tau$ of the electric dipole moment in the rest frame is indeed problematic, we look at the relation between $\pmb{d}$ and $\pmb{\mu}$ in the explicit form of Eq.~(\ref{no_EDM_rest_explicit}), which already takes into account $\pmb{d}_\text{rest}=0$. Taking the proper time derivative on both sides yields
\begin{equation}
\begin{aligned}
    \frac{\mathrm{d}\pmb{d}}{\mathrm{d}\tau}  &= \frac{\mathrm{d}}{\mathrm{d}\tau} \bigg( \frac{1}{c^2} \mathbf{v}\times \pmb{\mu} \bigg) \\
    &= \frac{1}{c^2} \mathbf{v}\times \bigg( \frac{\mathrm{d}\pmb{\mu}}{\mathrm{d}\tau}\bigg)  + \frac{1}{c^2} \bigg( \frac{\mathrm{d}\mathbf{v}}{\mathrm{d}\tau} \bigg) \times \pmb{\mu} \ .
\end{aligned}
\end{equation}
The form that the above equation takes in an inertial frame, in which the particle is instantaneously at rest, is
\begin{equation}
\label{ddot_rest}
    \bigg( \frac{\mathrm{d}\pmb{d}}{\mathrm{d}\tau}\bigg) \bigg|_\text{rest}=-\frac{e}{mc^2} (\pmb{\mu}\times\mathbf{E})|_\text{rest}\ ,
\end{equation}
where we have once again used $\mathbf{v}_\text{rest}=0$ to drop the first term on the right-hand side and the Lorentz force in an instantaneous rest frame (see the space component of Eq.~(\ref{lorentz_force_noncovariant}))
\begin{equation}
    \bigg( \frac{\mathrm{d}\mathbf{v}}{\mathrm{d}\tau} \bigg)\bigg|_\text{rest} = \frac{e}{m}\mathbf{E}_\text{rest}
\end{equation}
to reformulate the second term. We see that the second line of Eq.~(\ref{ddot_mudot_rest_incorrect}) is Eq.~(\ref{ddot_rest}) multiplied by $g/2$. The mismatch seems to be of significance, insofar as we consider Eq.~(\ref{ddot_rest}) to hold since it is derived from the Frenkel condition. There are seemingly different ways to go about fixing this problem in literature (see, for example, \cite{frenkel1926electrodynamics} or \cite{Bagrov_Classical_ST}). The most intuitive and elegant one that does not involve enforcing {\it a priori} restrictions is to add a zero term to the energy functional. This addition is a mathematical tool, and it lets us enforce conditions by introducing a new parameter, which we can then determine by utilizing the desired constraint, i.e., Eq.~(\ref{no_EDM_rest}). We add a zero term that keeps the structure of the energy functional intact -- a double index contraction of two antisymmetric second-rank tensors -- as
\begin{equation} \label{energy_functional}
	\begin{aligned}
            U &= - \frac{1}{2}\Sigma_{\mu\nu}F^{\mu\nu} \\ \rightarrow \tilde{U} &= - \frac{1}{2}\Sigma_{\mu\nu}(F^{\mu\nu} + \Xi^\mu u^\nu - \Xi^\nu u^\mu) \\ &= - \frac{1}{2}\Sigma_{\mu\nu}{\tilde{F}}^{\mu\nu} \ .
\end{aligned}
\end{equation}
We then calculate the equation of motion anew by substituting $F^{\mu\nu}$ with ${\tilde{F}}^{\mu\nu}$, which is now an effective field containing the auxiliary four-vector $\Xi^\mu$. The added terms in the energy functional vanish via the Frenkel condition, Eq.~(\ref{no_EDM_rest}). We do not make any assumptions about $\Xi^\mu$ and instead let its form emerge naturally. Such a derivation for an electron was partially done already by Frenkel \cite{frenkel1926electrodynamics}, but the missing knowledge at the time that the electron $g$-factor is, in fact, $g\neq 2$ yielded incomplete results. We take a general gyromagnetic ratio $g$ into account here. Replacing $F^{\mu\nu}$ by ${\tilde{F}}^{\mu\nu}$ in the commutator equation, Eq.~(\ref{eq:sigma_dot_commutator}) leads to
\begin{equation} \label{eq:commutatior_with_F'}
    \begin{aligned}
        \frac{\mathrm{d}\Sigma^{\mu\nu}}{\mathrm{d}\tau} &= \frac{eg}{2m} \big( F^\mu  _{\ \alpha}\Sigma^{\alpha\nu} - F^\nu  _{\ \alpha}\Sigma^{\alpha\mu} \big) \\ & \ \ \ \ - \frac{eg}{2m} \Xi_\alpha \big( u^\mu \Sigma ^{\alpha\nu} - u^\nu \Sigma ^{\alpha\mu} \big) \ 
    \end{aligned}
\end{equation}
and we can once again use the Frenkel condition to find the unknown four-vector $\Xi^\mu$
\begin{equation} \label{Xi_alpha}
    \Xi^\mu= \frac{1}{gc^2}(g-2)F^{\nu\mu}u_\nu \ .
\end{equation}
The detailed derivation can be found in Appendix~\ref{appendixC2_effLlikeTonMomentTensor}. We note that $\Xi^\mu$ is space-like, i.e., $\Xi^\mu u_\mu=0$. The evolution equation for the electromagnetic moment tensor hence becomes
\begin{equation}
\begin{aligned}
\label{sigma_dot}
    \frac{\mathrm{d}\Sigma^{\mu\nu}}{\mathrm{d}\tau} &= \frac{eg}{2m} \big( F^\mu  _{\ \alpha}\Sigma^{\alpha\nu} - F^\nu  _{\ \alpha}\Sigma^{\alpha\mu} \big) \\ & \ \ \ \ - \frac{e(g-2)}{2mc^2} \big(u^\mu\Sigma^{\nu\alpha} - u^\nu \Sigma^{\mu\alpha}\big)F_{\alpha\beta}u^\beta \ .
\end{aligned}
\end{equation}
The above equation agrees with the evolution equations in \cite{Bagrov_Classical_ST,Bordovitsyn_2001_SPINORBITAL}, which are derived in different ways than the one above. We have now found an additional term, which is proportional to $(g-2)$ and corresponds to the contribution of an anomalous magnetic moment that an electrically charged fermion might have. We know that the intrinsic magnetic moments of the electron and the muon do have an anomalous part \cite{pdg_amdm}. We see that in this classical treatment, the corresponding contribution arises naturally when we keep $g$ general and do not set it to $2$. As we see in the second term on the right-hand side of Eq.~(\ref{sigma_dot}), the evolution of the moment tensor is coupled to the four-velocity of the particle, i.e., its trajectory, through the electromagnetic field. This was not the case in Eq.~(\ref{eq:sigma_dot_commutator}), where $\Sigma ^{\mu\nu}$ evolves independently of $u^\mu$. For $g=2$ in Eq.~(\ref{sigma_dot}), we recover Eq.~(\ref{eq:sigma_dot_commutator}).

The space-time component of Eq.~(\ref{sigma_dot}) in a frame, which is momentarily at rest, indeed matches Eq.~(\ref{ddot_rest}) exactly. We have established an agreement between the evolution equation for the magnetic sources and the Frenkel condition, that is, the absence of an EDM in the rest frame. It turns out that we do not need the term $\Big( \frac{\mathrm{d}}{\mathrm{d}\tau}\pmb{d}\Big) \Big|_\text{rest}$ to vanish in order for the evolution equation to be correct. Instead, it is not identically zero and actually agrees with the relation resulting from Eqs.~(\ref{mu_d_transformation_props})--(\ref{no_EDM_rest}), i.e., Eq.~(\ref{ddot_rest}).

Furthermore, we take the effective electromagnetic field tensor ${\tilde{F}}^{\mu\nu}$ and write a Lorentz-like transformation with it in analogy to the case with just $F^{\mu\nu}$
\begin{equation}
\begin{aligned}
\label{eq:eff_Lorentz_like_transf}
    {\tilde{\Lambda}} ^\mu _{\ \nu} & = \delta ^\mu _\nu + \Delta\tau\frac{eg}{2m} \begin{pmatrix}
        0 & \tilde{E}_x /c & \tilde{E}_y /c & \tilde{E}_z /c \\
         \tilde{E}_x /c & 0 & \tilde{B}_z & -\tilde{B}_y \\
        \tilde{E}_y /c & -\tilde{B}_z & 0 & \tilde{B}_x \\
         \tilde{E}_z /c & \tilde{B}_y & -\tilde{B}_x & 0 
    \end{pmatrix} \\
    &= \delta ^\mu _\nu + \Delta\tau\frac{eg}{2m} {\tilde{F}}^\mu  _{\ \nu}
\end{aligned}
\end{equation}
with the effective fields explicitly given by
\begin{equation} \label{effective_fields_explicit}
    \begin{aligned}
        \tilde{\mathbf{E}} &= \mathbf{E} + \frac{(g-2)}{gc^2}\gamma ^2 \big( \mathbf{v} (\mathbf{v}\cdot\mathbf{E}) - c^2 (\mathbf{E}+\mathbf{v}\times\mathbf{B}) \big) \\
        \tilde{\mathbf{B}} &=\mathbf{B} - \frac{(g-2)}{gc^2}\gamma ^2 \big( \mathbf{v} (\mathbf{v}\cdot\mathbf{B})
        - v^2 \mathbf{B} + 
        \mathbf{v}\times\mathbf{E} \big)
    \end{aligned}
\end{equation}
and call it an effective Lorentz-like transformation. Analogously to ${\Lambda} ^\mu _{\ \nu}$, its inverse is given by
\begin{equation}
\label{inverse_of_effLlike_transformation}
    {\tilde{\Lambda}} _{\nu } ^{\ \mu} = ({\tilde{\Lambda}} ^{-1}) ^\mu  _{\ \nu} = \delta ^\mu _\nu - \Delta\tau\frac{eg}{2m} {\tilde{F}}^\mu  _{\ \nu} \ .
\end{equation}
and it keeps the metric tensor invariant. This can be readily seen from calculations equivalent to those in Appendix~\ref{appendixB2_inversetranformation} and~\ref{appendixB3_metrictensorinvariance} with the substitution $\frac{e}{m} {F}^\mu  _{\ \nu} \rightarrow \frac{eg}{2m} {\tilde{F}}^\mu  _{\ \nu}$. Transforming the electromagnetic moment tensor according to the transformation rules used in Eq.~(\ref{sigma'_transf}) and reconstructing the proper time derivative by utilizing the method that led to Eq.~(\ref{eq:sigma_dot_commutator}), now yields Eq.~(\ref{sigma_dot}). The effective Lorentz-like transformation can still be expressed in terms of the Lorentz boost and rotation generators, $\mathbf{K}$ and $\mathbf{J}$, respectively, as in Eq.~(\ref{active_infin_Lorentz_transformation}), with parameters
\begin{equation}
\label{effective_parameter_substitution}
        \pmb{\zeta} \rightarrow  \Delta\tau\frac{eg}{2m} \frac{\tilde{\mathbf{E}}}{c} \ , \
        \pmb{\varphi} \rightarrow -\Delta\tau\frac{eg}{2m} \tilde{\mathbf{B}} \ .
\end{equation}
with $\tilde{\mathbf{E}}$ and $\tilde{\mathbf{B}}$ given in Eq.~(\ref{effective_fields_explicit}).
Hence, we can keep the idea of the electromagnetic fields driving the sources by means of Lorentz boosts and rotations. In this case, we have a totally antisymmetric effective electromagnetic tensor ${\tilde{F}}^{\mu\nu}$, or adjusted boost and rotation parameters, which act as a coupler between the evolution of the magnetic sources and the trajectory of the particle.

The effective Lorentz-like transformation in Eq.~(\ref{eq:eff_Lorentz_like_transf}) originates from adjusting the Lorentz-like transformation acting on the electric sources, Eq.~(\ref{new_transformation}), to fit the properties and constraints of the evolution of magnetic sources, leading to Eq.~(\ref{sigma_dot}). However, we can show that when we let the effective Lorentz-like transformation act on the electric sources, that is, the four-velocity $u^\mu$, we reproduce the Lorentz force again. The reason why both $\frac{e}{m}{F}^\mu  _{\ \nu}$ and $\frac{eg}{2m}{\tilde{F}}^\mu  _{\ \nu}$ in Eqs.~(\ref{new_transformation}) and~(\ref{eq:eff_Lorentz_like_transf}), respectively, yield the correct result for the electric sources is the following. In the latter case the emerging terms proportional to $g$ cancel, while the remaining term yields the prefactor $\frac{e}{m}$ in the Lorentz force, which the former expression does directly. The explicit calculation is given in Appendix~\ref{appendixC3_effLlikeTonElectricSources}. As the arising differential equation remains the same, Eq.~(\ref{lorentz_force_covariant}), the exponentiation given in \cite[Chapter~1.1.3]{Itzykson_Zuber_QFT} is still its solution. Importantly, we can now use a single transformation, namely, the effective Lorentz-like one given in Eq.~(\ref{eq:eff_Lorentz_like_transf}), apply it to both fundamental sources -- electric and magnetic -- and correctly recreate their respective evolution equations in the presence of electromagnetic fields, which is a quite satisfactory unification. Hence, we can interpret the driving forces of both electric and magnetic sources as a series of infinitesimal boosts and rotations containing effective electromagnetic fields.

\subsection{Explicit Form and Non-Relativistic Limit}
\label{subsec:explicit_nonrel_magnetic_evolution}
The explicit form of the evolution equation for $\Sigma^{\mu\nu}$ in Eq.~(\ref{sigma_dot}) reads as
\begin{widetext}
\begin{equation} \label{eq:sigma_dot_explicit}
\begin{aligned} 
    \frac{\mathrm{d(\gamma\pmb{\mu})}}{\mathrm{d}\tau} & = \frac{eg}{2m}\gamma ( \pmb{d} \times \mathbf{E} + \pmb{\mu}\times \mathbf{B}) - \frac{e(g-2)}{2mc^2}\gamma^3  \Big( (\mathbf{v}\cdot\mathbf{E})(\mathbf{v}\times\pmb{d}) + \mathbf{v}\times(\pmb{\mu}\times\mathbf{E}) - (\mathbf{v}\cdot\pmb{\mu})(\mathbf{v}\times\mathbf{B}) \Big) \ , \\
    \frac{\mathrm{d(\gamma\pmb{d})}}{\mathrm{d}\tau} & = \frac{eg}{2m}\gamma\bigg( \pmb{d} \times \mathbf{B} - \frac{1}{c^2} \pmb{\mu}\times \mathbf{E}\bigg)- \frac{e(g-2)}{2mc^2}\gamma^3 \Big( \mathbf{E}\times (\mathbf{v}\times\pmb{d}) - \pmb{\mu}\times\mathbf{E} - \pmb{\mu}\times(\mathbf{v}\times\mathbf{B}) - \mathbf{v}\big( \mathbf{B}\cdot (\mathbf{v}\times\pmb{d}) \big) \Big) \ ,
\end{aligned}
\end{equation}
which can be reformulated to (see Appendix~\ref{appendixC4_explicitEqsforDipoleMoments} for details) 
\begin{equation}
    \label{eq:sigma_dot_explicit_subst}
    \begin{aligned}
        \gamma \frac{\mathrm{d}\pmb{\mu}}{\mathrm{d}t} &= \frac{eg}{2mc^2}\big( \pmb{\mu}\times\mathbf{B} - \mathbf{v}(\pmb{\mu}\cdot\mathbf{E}) \big) + \frac{e(g-2)}{2mc^2}\gamma ^2 (\mathbf{v}\cdot\pmb{\mu}) \bigg( \mathbf{E} + \mathbf{v}\times\mathbf{B} - \frac{1}{c^2} \mathbf{v}(\mathbf{v}\cdot\mathbf{E} ) \bigg) \ , \\
        \gamma \frac{\mathrm{d}\pmb{d}}{\mathrm{d}t} &= \frac{eg}{2mc^2} \mathbf{v}\times(\pmb{\mu}\times\mathbf{B}) -\frac{e}{mc^2} \pmb{\mu}\times \bigg( \mathbf{E} + \mathbf{v}\times\mathbf{B} - \frac{1}{c^2}\mathbf{v}(\mathbf{v}\cdot\mathbf{E}) \bigg) + \frac{e(g-2)}{2mc^4}\gamma ^2 ( \mathbf{v}\cdot\pmb{\mu} ) \mathbf{v}\times(\mathbf{E}+\mathbf{v}\times\mathbf{B}) \, ,
    \end{aligned}
\end{equation}
\end{widetext}
where now $\frac{\mathrm{d}}{\mathrm{d}t}$ appears instead of $\frac{\mathrm{d}}{\mathrm{d}\tau}$, while no non-relativistic approximation has yet been made. In Eq.~(\ref{eq:sigma_dot_explicit_subst}) we have substituted $\pmb{d}= \frac{1}{c^2}\mathbf{v}\times\pmb{\mu}$, the explicit form of Eq.~(\ref{no_EDM_rest}). Note that if the same substitution is made for $\frac{\mathrm{d}\pmb{d}}{\mathrm{d}t}$ in the second line in Eq.~(\ref{eq:sigma_dot_explicit_subst}), it can be shown that there is no new information in the evolution equation for the electric dipole moment. Hence, all of the information about the evolution of the system is contained in the first equation above, the equation for the magnetic moment, and the dynamics of the electric moment can be extracted from it via the explicit formulation of Eq.~(\ref{no_EDM_rest}). This finding is in agreement with the lack of an independent intrinsic electric dipole moment. 

Using the expressions in Eq.~(\ref{eq:sigma_dot_explicit_subst}), we can take the non-relativistic limit valid at small velocity values. The expansion of the above equation in powers of $\frac{v}{c}$, keeping terms up to and including second order, yields
\begin{widetext}
\begin{equation} \label{eq:sigma_dot_NR_O2}
\begin{aligned}
        \bigg( 1 + \frac{v^2}{2c^2}\bigg)\frac{\mathrm{d\pmb{\mu}}}{\mathrm{d}t} &= \frac{eg}{2m} \Bigg(   \pmb{\mu}\times\mathbf{B}  - \frac{1}{c}(\pmb{\mu} \cdot\mathbf{E} )\frac{\mathbf{v}}{c} \Bigg)  +\frac{e(g-2)}{2m}  \bigg(  \frac{\mathbf{v}}{c}\cdot \pmb{\mu} \bigg) \bigg( \frac{1}{c}\mathbf{E} + \frac{\mathbf{v}}{c}\times\mathbf{B} \bigg) \ , \\
        \bigg( 1 + \frac{v^2}{2c^2}\bigg)\frac{\mathrm{d\pmb{d}}}{\mathrm{d}t} &= \frac{eg}{2mc} \frac{\mathbf{v}}{c}\times(\pmb{\mu}\times\mathbf{B}) - \frac{e}{mc^2} \pmb{\mu}\times\Bigg( \mathbf{E}  + c \frac{\mathbf{v}}{c}\times\mathbf{B} - \frac{\mathbf{v}}{c} \bigg( \frac{\mathbf{v}}{c}\cdot\mathbf{E} \bigg) \Bigg)+\frac{e(g-2)}{2mc^2} \bigg( \frac{\mathbf{v}}{c} \cdot\pmb{\mu} \bigg) \frac{\mathbf{v}}{c} \times\mathbf{E} \ ,
 \end{aligned}
\end{equation}
\end{widetext}
where we have substituted the explicit form of Eq.~(\ref{no_EDM_rest}) for $\pmb{d}$. Making the non-relativistic approximation stricter and neglecting terms proportional to $\frac{v^2}{c^2}$ gives
\begin{equation} \label{eq:sigma_dot_NR_O1}
\begin{aligned}
        \frac{\mathrm{d\pmb{\mu}}}{\mathrm{d}t} &= \frac{eg}{2m} \bigg(  \pmb{\mu}\times\mathbf{B}  - \frac{1}{c}(\pmb{\mu} \cdot\mathbf{E} )\frac{\mathbf{v}}{c} \bigg) +\frac{e(g-2)}{2mc}  \bigg(  \frac{\mathbf{v}}{c}\cdot \pmb{\mu} \bigg)\mathbf{E} \ , \\
        \frac{\mathrm{d\pmb{d}}}{\mathrm{d}t} &= \frac{eg}{2mc} \frac{\mathbf{v}}{c}\times(\pmb{\mu}\times\mathbf{B}) - \frac{e}{mc^2} \pmb{\mu}\times\bigg( \mathbf{E}  + c \frac{\mathbf{v}}{c}\times\mathbf{B}\bigg) \ .
 \end{aligned}
\end{equation}
When the particle is at rest, $\frac{v}{c}=0$, the equation for $\pmb{\mu}$ coincides with the first line of Eq.~(\ref{ddot_mudot_rest_incorrect}), while that for $\pmb{d}$ collapses to Eq.~(\ref{ddot_rest}). Both equations correspond to our requirements for the physical system at rest, namely, no independent EDM and precession of the magnetic moment around the magnetic field.

\subsection{Spin-Pseudovector Representation}
\label{subsec:spin_pseudovector_rep}
For the purposes of completeness and relating this work further to existing literature, we introduce a four-pseudovector, $s^\mu$, which is also used to describe the classical spin of a particle. It is given by
\begin{equation}
    s^\mu = \begin{pmatrix}
        s_0 \\ \mathbf{s}
    \end{pmatrix} 
\end{equation}
with the temporal component vanishing in the instantaneous rest frame and the spatial ones being connected to the intrinsic magnetic moment via $\mathbf{s}_\text{rest} =  \frac{1}{\mu}\pmb{\mu}_\text{rest}$. Similarly to $\Sigma ^{\mu\nu}$, $s^\mu$ should also only have two degrees of freedom, but such a vector generally has four. One of them is eliminated by the condition $s^\mu s_\mu = -1 $. The other degree of freedom is removed by the Frenkel condition, which in this setting is formulated as \cite{tamm_1929,bmt_paper}
\begin{equation} \label{su=0}
    s^\mu u_\mu =0 \ ,
\end{equation}
from which the connection $s^0 =\frac{1}{c}\mathbf{v}\cdot\mathbf{s}$ arises. Applying the effective Lorentz-like transformation on the spin-pseudovector and reconstructing the proper time derivative using the same argument as before yields the evolution equation for $s^\mu$ (see Appendix~\ref{appendixC5_effLlikeTonSpinPseudovector} for details)
\begin{equation}
\label{eq:sdot}
    \frac{\mathrm{d}s ^\mu}{\mathrm{d}\tau} = \frac{eg}{2m}{F}^\mu  _{\ \nu} s^\nu - \frac{e (g-2)}{2mc^2} u^\mu F _{\nu \alpha} u^\nu  s^\alpha \ .
\end{equation}
The above equation matches the one in literature on the evolution of the magnetic moment in terms of a four-pseudovector (see, e.g., \cite{silenko_gamma_paper,Bagrov_Classical_ST,barut1980electrodynamics}). The two descriptions of classical spin -- the tensor $\Sigma^{\mu\nu}$ and vector $s^\mu$ ones -- are equivalent and a dual transformation exists between them \cite[Chapter 10.4]{wu-ki_Tung}, which reads as
\begin{equation} \label{dual_transf}
    s^\mu = \frac{1}{2\mu c}\epsilon ^{\mu\nu\alpha\beta}\Sigma_{\alpha\beta}u_\nu \ \ \ \text{and} \ \ \ \Sigma ^{\mu\nu} = \frac{\mu}{c} \epsilon ^{\mu\nu\alpha\beta} s_\alpha u_\beta \ .
\end{equation}
Equations~(\ref{sigma_dot}) and~(\ref{eq:sdot}) are compatible with the dual transformation in the above equation, and one can use it to switch between the evolution equations in the two representations. The corresponding calculation is sketched in Appendix~\ref{appendixD_dualtransfTensor_Pseudovector}.

In a final remark, we give the explicit connections between the separate components of the two classical spin representations $\Sigma^{\mu\nu}$ and $s^\mu$, which arise from Eq.~(\ref{dual_transf}). They read as
\begin{equation}
    \begin{gathered}
        s^0 = \frac{\gamma ^2}{\mu c} \mathbf{v}\cdot\pmb{\mu} \ , \ \ \mathbf{s} = \frac{1}{\mu}\bigg(\pmb{\mu}+\frac{\gamma ^2}{c^2}\mathbf{v}(\mathbf{v}\cdot\pmb{\mu})\bigg) \ , \\
        \pmb{\mu} = \mu \bigg( \mathbf{s} - \frac{1}{c^2} \mathbf{v}(\mathbf{v}\cdot\mathbf{s})\bigg) \ , \ \ \pmb{d} = \frac{\mu}{c^2} \mathbf{v}\times\mathbf{s} 
    \end{gathered}
\end{equation}
and are slightly different than the way they usually appear in literature, which is due to the additional factor $\gamma$ that we have in the definition of our tensor $\Sigma^{\mu\nu}$, which ensures the correct transformation properties of $\Sigma^{\mu\nu}$ upon a Lorentz boost.

\section{Conclusions}
\label{sec:conclusions}
In this study, we propose, investigate, and confirm a previously unexplored perspective on the driving mechanism of both kinds of fundamental electromagnetic sources in the presence of homogeneous fields. We show that we can interpret the evolution of point-like electric charges and magnetic moments as a series of infinitesimal effective Lorentz-like transformations acting on the corresponding sources. The transformations contain boosts and rotations with effective electric and magnetic fields as parameters, respectively. Intriguingly and importantly, we find that we can apply the exact same transformation on both the electric as well as the magnetic sources and correctly obtain their corresponding evolution equations. This discovery provides a consistent and unified interpretation of the forces driving the elementary electromagnetic sources. Such a union points to a fundamentally equivalent behavior of what is typically regarded as two inherently different intrinsic properties of particles -- electric charge and magnetic moment.

Understanding the fundamental behavior of particles with certain properties is not only satisfactory to the scientific curiosity intrinsic to scientists, but it can also lead to unpredictable progress. What steered us to the topic of the current work is the study of light-matter interaction in cases where magnetism plays a role. The underlying unit behind macroscopic magnetism is the spin of particles. Understanding its evolution and interpreting it suitably can help with translating the dynamics to a larger spatial scale. The perspective we give in this article could potentially be used to take the description of the interaction between electromagnetic fields and point-like sources to one between fields and continuous matter densities. Such an approach would be intuitive since Lorentz transformations acting on covariant elements such as four-vectors and second-rank tensors are well-understood.

A possible next step regarding the interpretation proposed in this study is to include contributions from inhomogeneous fields in the evolution equations for the sources and to explore the prospect of expressing the latter again as a series of infinitesimal Lorentz transformation with adjusted parameters. When the fields are not homogeneous, new terms in both evolution equations arise, and then there is not only an effect of the trajectory on the magnetic moment but also a back-coupling of the trajectory to the magnetic moment. In the latter case, the new terms represent a Stern-Gerlach-type force, whose covariant form is discussed in \cite{rafelski_covSG, formanek_covariantSG_charged_extPW,Formanek_covSG_neutral_extPW,Formanek_covSG_neutral_extlinearPW}.

\section*{Acknowledgments}
We acknowledge support by the KIT Publication Fund of the Karlsruhe Institute of Technology. I.I. acknowledges support by the Karlsruhe School of Optics \& Photonics (KSOP) and the Ministry of Science, Research and Arts of Baden-Württemberg as part of the sustainability financing of the projects of the Excellence Initiative II.

\appendix

\section{Equivalence of Eqs.~(\ref{no_EDM_rest_explicit}) and (\ref{no_EDM_rest})}

\label{appendixA_Eqs11_12}

We start with the first line of Eq.~(\ref{mu_d_transformation_props}) and apply $\mathbf{v}\times$ to it on the left to show that, using $\pmb{d}_\text{rest}=0$, 
\begin{equation}
    \mathbf{v}\times\pmb{\mu}_\text{rest} = \mathbf{v}\times\pmb{\mu} \ .
\end{equation}
The above result can be substituted into Eq.~(\ref{no_EDM_rest_explicit}) to yield 
\begin{equation} \label{eq:disvxmu}
    \pmb{d} = \frac{1}{c^2} \mathbf{v}\times \pmb{\mu} \ ,
\end{equation}
which in agreement with the space part of Eq.~(\ref{no_EDM_rest}), the explicit form of which reads as
\begin{equation}
    \Sigma^{\mu\nu}u_\nu = \gamma^2 \begin{pmatrix}
        -c\mathbf{v}\cdot\pmb{d} \\ -c^2\pmb{d} + \mathbf{v}\times\pmb{\mu}
    \end{pmatrix} = 0 \ .
\end{equation}
The temporal component then vanishes automatically, showing the equivalence of Eqs.~(\ref{no_EDM_rest_explicit}) and~(\ref{no_EDM_rest}).

\section{General Properties of the Lorentz and Lorentz-like Transformation}
\label{appendixB_generalproperties}
\subsection{Infinitesimal Lorentz Transformation}
\label{appendixB1_infLT}
Depending on convention, the infinitesimal Lorentz transformation consisting of a rotation and a boost can be represented differently. Here, we use the form given in Jackson, Section 11.7 \cite[Eqs.~(11.91) and (11.93)]{jackson1999classical}. The generators of rotation $J_i$ and those of Lorentz boosts $K_i$ for the three spatial axes read as
\begin{equation}
    \begin{aligned}
        J_x=\begin{pmatrix}
        0 & 0 & 0 & 0 \\
        0 & 0 & 0 & 0 \\
        0 & 0 & 0 & -1 \\
        0 & 0 & 1 & 0 \\
        \end{pmatrix} \ , \ \ 
        J_y=\begin{pmatrix}
        0 & 0 & 0 & 0 \\
        0 & 0 & 0 & 1 \\
        0 & 0 & 0 & 0 \\
        0 & -1 & 0 & 0 \\
        \end{pmatrix} \ , \\ 
        J_z=\begin{pmatrix}
        0 & 0 & 0 & 0 \\
        0 & 0 & -1 & 0 \\
        0 & 1 & 0 & 0 \\
        0 & 0 & 0 & 0 \\
        \end{pmatrix} \ , \ \
        K_x=\begin{pmatrix}
        0 & 1 & 0 & 0 \\
        1 & 0 & 0 & 0 \\
        0 & 0 & 0 & 0 \\
        0 & 0 & 0 & 0 \\
        \end{pmatrix} \ , \\ 
        K_y=\begin{pmatrix}
        0 & 0 & 1 & 0 \\
        0 & 0 & 0 & 0 \\
        1 & 0 & 0 & 0 \\
        0 & 0 & 0 & 0 \\
        \end{pmatrix} \ , \ \ 
        K_z=\begin{pmatrix}
        0 & 0 & 0 & 1 \\
        0 & 0 & 0 & 0 \\
        0 & 0 & 0 & 0 \\
        1 & 0 & 0 & 0 \\
        \end{pmatrix} \ . 
    \end{aligned}
\end{equation}
With these, the {\it active} Lorentz transformation reads as the first line in Eq.~(\ref{active_infin_Lorentz_transformation})
\begin{equation*}
    L ^\mu  _{\  \nu}  = \delta ^\mu _\nu + (\pmb{\varphi}\cdot \mathbf{J} + \pmb{\zeta}\cdot\mathbf{K})^\mu  _{\ \nu} \ .
\end{equation*}
The multiplication between the parameters and the generators represents a scalar product between the parameter vector and the vector of generator matrices, e.g., $\pmb{\varphi} = (\varphi_x,\varphi_y,\varphi_z)$ and $\mathbf{K}=(K_x,K_y,K_z)$, respectively. Performing the straightforward calculation yields the second line in Eq.~(\ref{active_infin_Lorentz_transformation})
\begin{equation*}
     L ^\mu  _{\ \nu}= \delta ^\mu _\nu + \begin{pmatrix}
        0 & \zeta _1 & \zeta _2 & \zeta _3 \\
        \zeta _1 & 0 & -\varphi _3 & \varphi _2 \\
        \zeta _2 & \varphi _3 & 0 & -\varphi _1 \\
        \zeta _3 & -\varphi _2 & \varphi _1 & 0 \\
    \end{pmatrix} 
\end{equation*}

\subsection{Inverse Transformation}
\label{appendixB2_inversetranformation}
Every infinitesimal transformation consists of an identity operation, here $\delta^\mu _\nu$, to which one adds the product of a small parameter and a generator, which describes the infinitesimal change. The inverse of that transformation is the same, only with a minus sign in front of the second term. Physically, the sign change represents the undoing of the small addition to the initial quantity. The product of such a transformation and its inverse yields an identity matrix up to first order in the small transformation parameter, and all higher orders are negligible since we consider an infinitesimal transformation. Hence, the inverse of the active Lorentz transformation in Eq.~(\ref{active_infin_Lorentz_transformation}) is given by\footnote{Depending on the definition, the transformation might include subtraction of the infinitesimal change instead of addition. The inverse will always have the corresponding opposite sign.}
\begin{equation}
    (L^{-1}) ^\mu  _{\  \nu}  = \delta ^\mu _\nu - (\pmb{\varphi}\cdot \mathbf{J} + \pmb{\zeta}\cdot\mathbf{K})^\mu  _{\ \nu} \ .
\end{equation}
The inverse of the Lorentz-like transformation driving the Maxwell sources to evolve emerges upon the parameter substitution in Eq.~(\ref{parameter_substitution}) and is given by Eq.~(\ref{inverse_of_new_transformation})
\begin{equation*}
        (\Lambda ^{-1}) ^\mu  _{\  \nu} = \delta ^\mu _\nu - \Delta\tau\frac{e}{m} F^\mu  _{\ \nu} \ .
\end{equation*}
Taking the new transformation $\Lambda ^\mu  _{\  \nu}$ and explicitly calculating its inverse up to first order, which is valid in the infinitesimal case, yields the same result. For the sake of completeness, we shall show that Eq.~(\ref{inverse_of_new_transformation}) is indeed the correct inverse
\begin{equation}
    \begin{aligned}
        \Lambda ^\mu  _{\  \beta} (\Lambda ^{-1}) ^\beta  _{\  \nu} &= \Big( \delta ^\mu _\beta + \Delta\tau\frac{e}{m} F^\mu  _{\  \beta} \Big) \Big(  \delta ^\beta _\nu - \Delta\tau\frac{e}{m} F^\beta  _{\  \nu} \Big) \\
        &= \delta ^\mu _\beta \delta ^\beta _\nu + \Delta\tau\frac{e}{m} \delta ^\beta _\nu F^\mu  _{\  \beta} - \Delta\tau\frac{e}{m} \delta ^\mu _\beta F^\beta  _{\  \nu} \\
        &= \delta^\mu _\nu +\Delta\tau\frac{e}{m} F ^\mu _{\  \nu} - \Delta\tau\frac{e}{m} F ^\mu _{\  \nu} \\
        &= \delta^\mu _\nu \ .
    \end{aligned}
\end{equation}
up to first order. In the first line above, we have substituted the transformation and its inverse with a fitting choice of index placement. In the second line, the parentheses have been expanded and any $\mathcal{O}(\Delta\tau^2)$-terms neglected, and the terms were simplified in the next line. Finally, we arrive at the last line after canceling the two identical terms with opposite signs. We get an identity matrix, $\delta^\mu _\nu$, proving that $(\Lambda ^{-1}) ^\mu  _{\  \nu}$ from Eq.~(\ref{inverse_of_new_transformation}) is in fact the inverse of $\Lambda ^\mu  _{\  \nu}$ from Eq.~(\ref{new_transformation}).

In the case of mixed tensors such as $\Lambda ^\mu  _{\ \nu}$, there are different conventions regarding the order of the indices and their character. For example, one may define $\Lambda ^\mu  _{\ \nu}$, a mixed tensor with its first index being contravariant and its second one -- covariant, to be the inverse of the tensor with the opposite order of its indices, i.e., $\Lambda ^\mu  _{\ \nu} \equiv (\Lambda ^{-1}) _\nu ^{\ \mu}$. While remaining compatible with this method, we shall show in the following how the index placement influences the transformation. Since the identity operation $\delta ^\mu _\nu$ is a symmetric matrix, we need not distinguish between the first and second index, that is, $\delta ^\mu _\nu = \delta ^\mu _{\ \nu} = \delta _\nu ^{\ \mu} $. We examine the electromagnetic tensor\footnote{This holds for every totally antisymmetric second-rank tensor.} and get
\begin{equation}
    \begin{aligned}
        F^\mu _{\  \nu} &= \eta ^{\mu\beta}F_{\beta\nu} \\
        &= -\eta ^{\mu\beta}F_{\nu\beta} \\
        &= -\eta ^{\beta\mu}F_{\nu\beta} \\
        &= -F_\nu ^{\  \mu} \ .
    \end{aligned}
\end{equation}
In the above calculations, several qualities of the two tensors are utilized. The first line shows how one gets $F^\mu _{\ \nu}$ from the covariant tensor $F_{\beta\nu}$. In the second row, the antisymmetric nature of the electromagnetic tensor is used, $F_{\beta\nu}=-F_{\nu\beta}$, while in the next one, the indices of the metric tensor are switched, which leads to no change in sign since it is symmetric. Finally, in the last line, the electromagnetic tensor with a first covariant and a second contravariant index is built, and one may notice the remaining minus sign in front of it, ultimately yielding the connection $F^\mu _{\ \nu} = -F_{\nu } ^{\  \mu}$. We now go back to the whole transformation and write
\begin{equation}
\label{L_to_L-1_relation}
    \begin{aligned}
        \Lambda ^\mu  _{\ \nu} &= \delta ^\mu _\nu + \Delta\tau\frac{e}{m} F^\mu  _{\ \nu} \\
        &= \delta ^\mu _\nu - \Delta\tau\frac{e}{m} F_\nu ^{\  \mu} \\
        &=(\Lambda ^{-1}) _\nu ^{\ \mu} \ ,
    \end{aligned}
\end{equation}
showing the previously {\it a priori} defined relation.

\subsection{Metric Tensor Invariance}
\label{appendixB3_metrictensorinvariance}
The cornerstone of special relativity is the invariance of the infinitesimal four-volume element, Eq.~(\ref{invariant_volume_element}). To assure this condition is fulfilled, the metric tensor $\eta _{\mu\nu}$ must also remain invariant. We can see this from the form of the transformed volume element by requiring that it be conserved
\begin{equation*}
    \begin{aligned}
        \mathrm{d}x^\mu \mathrm{d}x_\mu &= \eta ^{\mu\nu} \mathrm{d}x_\mu \mathrm{d}x_\nu \\
        &\stackrel{!}{=}  \eta ^{\mu\nu} \mathrm{d}x^{\prime}_{\mu} \mathrm{d}x^{\prime}_{\nu} \\
        &= \eta ^{\mu\nu} \big( \Lambda ^\beta  _{\  \mu} \Lambda ^\sigma  _{\  \nu}\mathrm{d}x_\beta \mathrm{d}x_\sigma \big) \\
        &= \big( \Lambda ^\beta  _{\  \mu} \Lambda ^\sigma  _{\  \nu} \eta ^{\mu\nu} \big) \mathrm{d}x_\beta \mathrm{d}x_\sigma \\
        &= \big( \Lambda ^\mu  _{\  \beta} \Lambda ^\nu  _{\  \sigma} \eta ^{\beta\sigma} \big) \mathrm{d}x_\mu \mathrm{d}x_\nu \ .
    \end{aligned}
\end{equation*}
In the first line of the above calculation, we have rewritten the volume element as two covariant elements, where one of them is contracted with the covariant metric tensor. In the next row, we apply the requirement that the four-volume element remains invariant and, hence, the expression must be equal to when the two infinitesimal line elements are first transformed and then contracted with the metric tensor, which is written explicitly in the third line. In the next one, we have regrouped the terms, which we are allowed to do since the index placement fully determines the multiplication. In this order, we can interpret the transformation matrices as acting on the metric tensor, even though we have done nothing else but shuffle the terms. Finally, in the last row, we renamed all of the dummy indices to make the comparison more obvious. Since we required that lines one and five are equivalent, we can formulate the invariance condition for the transformation of the metric tensor as
\begin{equation}
        \eta ^{\mu\nu} = \Lambda ^\mu  _{\  \beta} \Lambda ^\nu  _{\  \sigma} \eta ^{\beta\sigma} \ .
\end{equation}
To show this for our Lorentz-like transformation matrix $\Lambda$, we perform the calculation of the right-hand side of the above expression, which yields
\begin{equation}
    \begin{aligned}
        \Lambda ^\mu  _{\  \beta} \Lambda ^\nu  _{\  \sigma} \eta ^{\beta\sigma} &= \Lambda ^\mu  _{\  \beta} (\Lambda ^{-1})  _{\sigma} ^{\  \nu}  \eta ^{\beta\sigma}  \\
        &= \big( \delta ^\mu _\beta + \Delta\tau\frac{e}{m} F^\mu  _{\  \beta} \big) \big( \delta ^\nu _\sigma - \Delta\tau\frac{e}{m} F _{\sigma}^{\  \nu} \big) \eta ^{\beta\sigma} \\
        &=  \big( \delta ^\mu _\beta + \Delta\tau\frac{e}{m} F^\mu  _{\  \beta} \big) \big( \eta ^{\beta\nu} - \Delta\tau\frac{e}{m} F^{\beta\nu} \big) \\
        &=\eta ^{\mu\nu} - \Delta\tau\frac{e}{m} F^{\mu\nu} +  \Delta\tau\frac{e}{m} F^{\mu\nu} \\
        &= \eta ^{\mu\nu} \ .
    \end{aligned}
\end{equation}
In the above proof, we have first used the identity of the inverse from Eq.~(\ref{L_to_L-1_relation}) and have then written both transformations explicitly in the second line. In the next one, the terms in the second parentheses have been applied on $\eta ^{\beta\sigma}$. In the fourth row, the two sets of parentheses have been expanded again up to first order, resulting in two terms, which cancel exactly to give the contravariant metric tensor we needed to fulfill invariance under the Lorentz-like transformation.

\section{Lorentz-like Transformations for the Evolution of the Electromagnetic Sources}
\label{appendixC_LlikeTforEMsources}
\subsection{Lorentz-like Transformation Acting on the Moment Tensor}
\label{appendixC1_LlikeTonMomentTensor}
In the following, we give a more detailed version of the calculation in Eq.~(\ref{sigma'_transf})
\begin{equation}
\begin{split}
    {\Sigma'} ^{\mu\nu} &= \Lambda ^\mu _{\ \alpha}
    \Lambda ^\nu _{\ \beta}  \Sigma^{\alpha\beta}   \\
    &= \big(\delta ^\mu _\alpha + \Delta\tau\kappa F^\mu  _{\ \alpha}\big)\big(\delta ^\nu _\beta + \Delta\tau\kappa F^\nu  _{\ \beta}\big)\Sigma^{\alpha\beta}\\
    &= \big(\delta ^\mu _\alpha + \Delta\tau\kappa F^\mu  _{\ \alpha}\big)\big(\Sigma^{\alpha\nu} + \Delta\tau\kappa F^\nu  _{\ \beta}\Sigma^{\alpha\beta} \big) \\
    &= \Sigma^{\mu\nu} + \Delta\tau\kappa F^\mu  _{\ \alpha}\Sigma^{\alpha\nu} + \Delta\tau\kappa F^\nu  _{\ \beta}\Sigma^{\mu\beta}\\
    &= \Sigma^{\mu\nu} + \Delta\tau\kappa \big( F^\mu  _{\ \alpha}\Sigma^{\alpha\nu} - F^\nu  _{\ \alpha}\Sigma^{\alpha\mu} \big) \ .
\end{split}
\end{equation}
The first line of the above equation represents the transformation law for second-rank tensors. In the next row, the two $\Lambda$'s have been expanded as given in Eq.~(\ref{new_transformation}) with a generalized constant $\kappa$. In the next two lines, the operations in the two sets of parentheses have been applied consecutively, leaving the three terms in the fourth line, where higher order terms $\mathcal{O}(\Delta\tau^2)$ have been dropped out since they are negligible in the infinitesimal regime. In the last row, the two terms proportional to $\Delta\tau\kappa$ have been grouped together. The indices of the tensor $\Sigma$ have been interchanged in the last term, giving rise to a change in sign, and the dummy index $\beta$ in that same term has been changed to $\alpha$ for later convenience.

\subsection{Effective Lorentz-like Transformation Acting on the Moment Tensor}
\label{appendixC2_effLlikeTonMomentTensor}
Replacing $F^{\mu\nu}$ by ${\tilde{F}}^{\mu\nu}$ in the commutator equation for the electromagnetic moment tensor, Eq.~(\ref{eq:sigma_dot_commutator}) yields
\begin{widetext}
\begin{equation} \label{eq:commutatior_with_F'_Appendix}
    \begin{aligned}
        \frac{\mathrm{d}\Sigma^{\mu\nu}}{\mathrm{d}\tau}  &= \frac{eg}{2m} \big( {\tilde{F}}^\mu  _{\ \alpha}\Sigma^{\alpha\nu} - {\tilde{F}}^\nu  _{\ \alpha}\Sigma^{\alpha\mu} \big) \\ 
        &= \frac{eg}{2m} \big( F^\mu  _{\ \alpha}\Sigma^{\alpha\nu} - F^\nu  _{\ \alpha}\Sigma^{\alpha\mu} \big) +  \frac{eg}{2m} u_\alpha \big( \Sigma ^{\alpha\nu}\Xi^\mu - \Sigma ^{\alpha\mu}\Xi^\nu \big) - \frac{eg}{2m} \Xi_\alpha \big( u^\mu \Sigma ^{\alpha\nu} - u^\nu \Sigma ^{\alpha\mu} \big) \\
        &= \frac{eg}{2m} \big( F^\mu  _{\ \alpha}\Sigma^{\alpha\nu} - F^\nu  _{\ \alpha}\Sigma^{\alpha\mu} \big) - \frac{eg}{2m} \Xi_\alpha \big( u^\mu \Sigma ^{\alpha\nu} - u^\nu \Sigma ^{\alpha\mu} \big) \ ,
    \end{aligned}
\end{equation}
\end{widetext}
where we have used that the last term in the second line vanishes as a consequence of contracting one of the indices of the moment tensor with the four-velocity, Eq.~(\ref{no_EDM_rest}). Taking the proper time derivative on both sides of the same condition, Eq.~(\ref{no_EDM_rest}), yields
\begin{equation} \label{sigmau_dot=0}
    \bigg(  \frac{\mathrm{d}\Sigma^{\mu\nu}}{\mathrm{d}\tau}   \bigg) u_\nu + \Sigma^{\mu\nu} \bigg( \frac{\mathrm{d}u_\nu}{\mathrm{d}\tau}  \bigg) =0 \ .
\end{equation}
We can use Eq.~(\ref{eq:commutatior_with_F'_Appendix}) and contracting all terms with $u_\nu$ to get the first term on the left-hand side of the above equation
\begin{equation} \label{sigmadotu}
    \begin{aligned}
        \bigg(  \frac{\mathrm{d}\Sigma^{\mu\nu} }{\mathrm{d}\tau}  \bigg) u_\nu &= \frac{eg}{2m} \big( F^\mu  _{\ \alpha}\Sigma^{\alpha\nu}u_\nu - F^\nu  _{\ \alpha}u_\nu\Sigma^{\alpha\mu} \big) \\ & \ \ \ \, - \frac{eg}{2m} \Xi_\alpha \big( u^\mu \Sigma ^{\alpha\nu} u_\nu - u^\nu u_\nu \Sigma ^{\alpha\mu} \big) \\
         &= - \frac{eg}{2m} F^\nu  _{\ \alpha}u_\nu\Sigma^{\alpha\mu} + \frac{egc^2}{2m} \Xi_\alpha  \Sigma ^{\alpha\mu} \ .
    \end{aligned}
\end{equation}
Above, we have used the Frenkel condition to drop the first terms in both sets of parentheses as well as the Lorentz invariant quantity $u^\mu u_\mu = c^2$. On the other hand, we have
\begin{equation} \label{sigmaudot}
    \begin{aligned}
         \Sigma^{\mu\nu} \bigg( \frac{\mathrm{d} u_\nu}{\mathrm{d}\tau} \bigg)
        &=  \frac{e}{m} \Sigma^{\mu\nu} F_{\nu\alpha} u^\alpha  \\
        &= - \frac{e}{m} \Sigma^{\alpha\mu} F_{\alpha\nu} u^\nu \ ,
    \end{aligned}
\end{equation}
where we have substituted the Lorentz force, Eq.~(\ref{lorentz_force_covariant}), in the first line and used $\Sigma^{\mu\nu}= -\Sigma^{\nu\mu}$ and then switched the dummy indices $\alpha$ and $\nu$ in the second one. \Cref{sigmaudot,sigmadotu,sigmau_dot=0} can then be combined into
\begin{equation}
    \frac{e}{m} \Sigma^{\alpha\mu}  \bigg( - \frac{g}{2} F^\nu  _{\ \alpha}u_\nu + \frac{gc^2}{2} \Xi_\alpha +F^\nu  _{\ \alpha}u_\nu \bigg) =0 \ ,
\end{equation}
by using that $F _{\alpha\nu }=-F _{\nu\alpha }$ and $ F_{\nu\alpha} u^\nu = F^\nu  _{\ \alpha}u_\nu $ to reformulate the right-hand side of Eq.~(\ref{sigmaudot}). Requiring that the whole term in parenthesis in the above equation vanishes lets us determine the vector $\Xi_\alpha$ to be
\begin{equation}
    \Xi_\alpha= \frac{1}{gc^2}(g-2)F^\nu  _{\ \alpha}u_\nu \ ,
\end{equation}
which corresponds to Eq.~(\ref{Xi_alpha}). Substituting the vector $\Xi_\alpha$ in Eq.~(\ref{eq:commutatior_with_F'_Appendix}) and rewriting $(u^\mu\Sigma^{\alpha\nu} - u^\nu\Sigma^{\alpha\mu})F^\nu _{\ \alpha}u_\nu = (u^\mu\Sigma^{\nu\alpha} - u^\nu\Sigma^{\mu\alpha})F_{\alpha\nu}u^\nu$ yields Eq.~(\ref{sigma_dot}).

\subsection{Effective Lorentz-like Transformation Acting on the Electric Sources}
\label{appendixC3_effLlikeTonElectricSources}
We apply the effective Lorentz-like transformation, Eq.~(\ref{eq:eff_Lorentz_like_transf}), on the four-velocity $u^\mu$. The calculation goes as follows
\begin{equation}
    \begin{aligned}
        {u'}^\mu &= {\tilde{\Lambda}}^\mu _{\ \nu} u^\nu \\
        &= \bigg( \delta ^\mu _\nu + \Delta\tau\frac{eg}{2m} {\tilde{F}}^\mu  _{\ \nu} \bigg) u^\nu \\
    &= \bigg( \delta ^\mu _\nu + \Delta\tau\frac{eg}{2m} \big( F^\mu  _{\ \nu} + \Xi^\mu u_\nu - \Xi_\nu u^\mu \big) \bigg) u^\nu \\
        &= \bigg(  \delta ^\mu _\nu + \Delta\tau\frac{eg}{2m} {F}^\mu  _{\ \nu}  + \Delta\tau\frac{e(g-2)}{2mc^2} F^{\alpha\mu}  u_\alpha u_\nu \\ & \ \ \ \, - \Delta\tau\frac{e(g-2)}{2mc^2} F  _{\alpha \nu} u^\alpha u^\mu \bigg)u^\nu \\
        &= u^\mu + \Delta\tau\frac{eg}{2m} {F}^\mu  _{\ \nu} u^\nu - \Delta\tau\frac{e(g-2)}{2m} F^{\mu}_{\ \alpha}  u^\alpha \\
        &= u^\mu + \Delta\tau\frac{e}{m} {F}^\mu  _{\ \nu} u^\nu \ ,
    \end{aligned}
\end{equation}
where we have applied the transformation on the four-velocity, which is written first in terms of ${\tilde{F}}^{\mu\nu}$, then in terms of $\Xi^\mu$ and finally in its most explicit but still covariant form in the first four steps of the above equation. Opening the parenthesis in the next line leads to the vanishing of the last term since $F  _{\alpha \nu} u^\alpha u^\nu = 0$. The property is easily obtained when one considers that on the one hand $F_{\alpha\nu} u^\alpha u^\nu = F_{\nu\alpha} u^\nu u^\alpha$ when simply exchanging the dummy indices $\alpha$ and $\nu$ and on the other hand $F_{\alpha\nu} u^\alpha u^\nu =- F_{\nu\alpha} u^\alpha u^\nu$ upon utilizing the total asymmetry of the electromagnetic tensor. Equating both results yields $F_{\nu\alpha} u^\nu u^\alpha = - F_{\nu\alpha} u^\alpha u^\nu$, which means that the expression can only be identically zero. In the sixth line we have also used that $u_\mu u^\mu = c^2$ in the third term as well as the reformulation $F^{\alpha\mu}  u_\alpha = - F^{\mu\alpha}  u_\alpha =- F^{\mu}_{\ \alpha}  u^\alpha$. Finally, we see that the second term and the part of the third one, which is proportional to $g$, are identical up to a sign and hence cancel. We are left with the result in the last line, which yields the Lorentz force directly upon the use of Eq.~(\ref{identification_Deltau_isu'}) and reconstruction of the proper time derivative. In short, it holds that $\frac{e}{m}{F}^\mu  _{\ \nu} u^\nu = \frac{eg}{2m}{\tilde{F}}^\mu  _{\ \nu} u^\nu$.

\subsection{Explicit Evolution Equations for the Electric and Magnetic Dipole Moments}
\label{appendixC4_explicitEqsforDipoleMoments}
We take the expressions in Eq.~(\ref{eq:sigma_dot_explicit}) and expand the left-hand sides as
\begin{equation}\label{dgammamu}
    \frac{\mathrm{d}(\gamma\pmb{\mu})}{\mathrm{d}\tau}=\frac{\mathrm{d}\gamma}{\mathrm{d}\tau} \pmb{\mu} + \gamma \frac{\mathrm{d}\pmb{\mu}}{\mathrm{d}\tau}
\end{equation}
and equivalently for $\gamma\pmb{d}$. We can now eliminate the term $\frac{\mathrm{d}\gamma}{\mathrm{d}\tau}$ by performing the proper time derivative on the definition of $\gamma$
\begin{equation}
\label{dgamma}
    \frac{\mathrm{d}\gamma}{\mathrm{d}\tau} = \frac{\mathrm{d}}{\mathrm{d}\tau} \frac{1}{\sqrt{1-\frac{\sqrt{\mathbf{v}\cdot\mathbf{v}}}{c^2}}} = \frac{\gamma^3}{c^2}\mathbf{v}\cdot\frac{\mathrm{d}\mathbf{v}}{\mathrm{d}\tau}
\end{equation}
and utilizing the space-part of the Lorentz force, Eq.~(\ref{lorentz_force_noncovariant}), expanding the derivative using the product rule as
\begin{equation}
\begin{aligned}
    \frac{\mathrm{d}(\gamma\mathbf{v})}{\mathrm{d}\tau} &= \mathbf{v}\frac{\mathrm{d}\gamma}{\mathrm{d}\tau} + \gamma\frac{\mathrm{d}\mathbf{v}}{\mathrm{d}\tau}  \\
    &=  \frac{\gamma^3}{c^2}\mathbf{v} \bigg(\mathbf{v}\cdot\frac{\mathrm{d}\mathbf{v}}{\mathrm{d}\tau} \bigg) + \gamma\frac{\mathrm{d}\mathbf{v}}{\mathrm{d}\tau}  \\
    &= \frac{e}{m}\gamma (\mathbf{E}+\mathbf{v}\times\mathbf{B}) \ ,
\end{aligned}
\end{equation}
where in the second line the result from Eq.~(\ref{dgamma}) has been substituted. Taking the last two lines of the above equation and projecting them onto $\mathbf{v}$ yields
\begin{equation}
    \begin{aligned}
        \gamma^3\frac{\mathbf{v}\cdot\mathbf{v}}{c^2} \bigg(\mathbf{v}\cdot\frac{\mathrm{d}\mathbf{v}}{\mathrm{d}\tau} \bigg) + \gamma\mathbf{v}\cdot\frac{\mathrm{d}\mathbf{v}}{\mathrm{d}\tau} &= \frac{e}{m}\gamma \mathbf{v}\cdot(\mathbf{E}+\mathbf{v}\times\mathbf{B}) \ , \\
        \bigg( \gamma^3\frac{v^2}{c^2} + \gamma \bigg)\bigg(\mathbf{v}\cdot\frac{\mathrm{d}\mathbf{v}}{\mathrm{d}\tau} \bigg) &= \frac{e}{m}\gamma \mathbf{v}\cdot\mathbf{E} \ , \\
        \mathbf{v}\cdot\frac{\mathrm{d}\mathbf{v}}{\mathrm{d}\tau} &= \frac{e}{m\gamma^{2}} \mathbf{v}\cdot\mathbf{E} \ ,
    \end{aligned}
\end{equation}
which we can then substitute in Eq.~(\ref{dgamma}) to get
\begin{equation} \label{dgamma_final}
\frac{\mathrm{d}\gamma}{\mathrm{d}\tau} = \frac{e}{mc^2}\gamma\mathbf{v}\cdot\mathbf{E} \ . 
\end{equation}
Plugging in Eqs.~(\ref{dgammamu}) and (\ref{dgamma_final}) into Eq.~(\ref{eq:sigma_dot_explicit}), using the identities $\mathbf{a}\cdot(\mathbf{b}\times\mathbf{c})= \mathbf{b}\cdot(\mathbf{c}\times\mathbf{a})$ and $\mathbf{a}(\mathbf{a}\cdot(\mathbf{b}\times\mathbf{c}))=(\mathbf{a}\times\mathbf{b})\times(\mathbf{a}\times\mathbf{c})$ to rewrite $\mathbf{v}(\mathbf{B}\cdot(\mathbf{v}\times\pmb{d})) = (\mathbf{v}\times\pmb{d})\times(\mathbf{v}\times\mathbf{B})$ in the evolution equation for $\pmb{d}$, substituting $\pmb{d}=\frac{1}{c^2}\mathbf{v}\times\pmb{\mu}$, and switching from proper to coordinate time with $\mathrm{d}\tau = \gamma^{-1}\mathrm{d}t$ results in
\begin{widetext}
\begin{equation}
    \begin{aligned}
         \gamma ^2 \frac{\mathrm{d}\pmb{\mu}}{\mathrm{d}t} &= \frac{eg}{2m}\gamma \bigg( \frac{1}{c^2}(\mathbf{v}\times\pmb{\mu})\times\mathbf{E} + \pmb{\mu}\times\mathbf{B} \bigg)  - \frac{e}{mc^2}\gamma(\mathbf{v}\cdot\mathbf{E})\pmb{\mu} \\ & \ \ \ - \frac{e(g-2)}{2mc^2}\gamma ^3\bigg( \frac{1}{c^2}(\mathbf{v}\cdot\mathbf{E})\big( \mathbf{v}\times(\mathbf{v}\times\pmb{\mu})\big) + \mathbf{v}\times(\pmb{\mu}\times\mathbf{E}) - (\mathbf{v}\cdot\pmb{\mu})(\mathbf{v}\times\mathbf{B})) \bigg) \ , \\
         \gamma ^2 \frac{\mathrm{d}\pmb{d}}{\mathrm{d}t} &= \frac{eg}{2mc^2}\gamma \big( (\mathbf{v}\times\pmb{\mu})\times\mathbf{B} - \pmb{\mu}\times\mathbf{E}\big) - \frac{e}{mc^4}\gamma(\mathbf{v}\cdot\mathbf{E})(\mathbf{v}\times\pmb{\mu}) \\ & \ \ \ - \frac{e(g-2)}{2mc^2}\gamma ^3\bigg( \frac{1}{c^2}\mathbf{E}\times\big( \mathbf{v}\times(\mathbf{v}\times\pmb{\mu}) \big) - \pmb{\mu}\times\mathbf{E} - \pmb{\mu}\times (\mathbf{v}\times\mathbf{B}) - \frac{1}{c^2} \big( \mathbf{v}\times (\mathbf{v}\times\pmb{\mu}) \big) \times(\mathbf{v}\times\mathbf{B})\bigg) \ .
    \end{aligned}
\end{equation}    
\end{widetext}
Application of the triple cross product identity $\mathbf{a}\times(\mathbf{b}\times\mathbf{c}) = (\mathbf{a}\cdot\mathbf{c})\mathbf{b} - (\mathbf{a}\cdot\mathbf{b})\mathbf{c}$ to multiple terms in the above equations, results in different but equivalent reformulations of the expressions, depending on the choice of term grouping. One possibility is Eq.~(\ref{eq:sigma_dot_explicit_subst}), which is a somewhat more compact formulation.

\subsection{Effective Lorentz-like Transformation Acting on the Spin-Pseudovector}
\label{appendixC5_effLlikeTonSpinPseudovector}
Applying the effective Lorentz-like transformation on the spin-pseudovector according to the transformation rule for a four-vector and writing down the transformation in terms of ${\tilde{F}^\mu _{\ \nu}}$ yields
\begin{equation}
\begin{aligned}
    {s'} ^\mu &= {\tilde{\Lambda}} ^\mu _{\ \nu} s^\nu \\
    &= \bigg( \delta ^\mu _\nu + \Delta\tau\frac{eg}{2m} {\tilde{F}}^\mu  _{\ \nu} \bigg) s^\nu \\
    &= \bigg( \delta ^\mu _\nu + \Delta\tau\frac{eg}{2m} \big( F^\mu  _{\ \nu} + \Xi^\mu u_\nu - \Xi_\nu u^\mu \big) \bigg) s^\nu \\
     &= \bigg( \delta ^\mu _\nu + \Delta\tau\frac{eg}{2m} {F}^\mu  _{\ \nu}  + \Delta\tau\frac{e(g-2)}{2mc^2} F^{\alpha\mu}  u_\alpha u_\nu \\ & \ \ \ - \Delta\tau\frac{e(g-2)}{2mc^2} F  _{\alpha \nu} u^\alpha u^\mu \bigg) s^\nu \\
     &= s^\mu \\ & \ \ \ + \Delta\tau \bigg( \frac{eg}{2m} {F}^\mu  _{\ \nu} s^\nu - \frac{e(g-2)}{2mc^2}  u^\mu F  _{\alpha \nu} u^\alpha  s^\nu \bigg) \ .
\end{aligned}
\end{equation}
Here, the third line $\tilde{F}^\mu _{\ \nu}$ has been expanded in terms of $\Xi^\mu$, whose form has been written explicitly in the next step. Finally, the condition in Eq.~(\ref{su=0}) has been used to drop the third term in the parentheses, resulting in the final form. 
Reconstructing the proper time derivative by taking ${s'}^\mu \rightarrow s^\mu (\tau +\Delta\tau)$ then leads to Eq.~(\ref{eq:sdot})

\section{Dual Transformation Between the Moment Tensor and Pseudovector}
\label{appendixD_dualtransfTensor_Pseudovector}
In order to perform the conversion between the moment tensor and pseudovector representation of classical spin, we need some contractions of four-dimensional Levi-Civita symbols given by
\begin{subequations}
    \begin{align}
        \begin{split}
        \epsilon_{\mu\nu\alpha\beta}\epsilon^{\mu\gamma\delta\sigma} &= -\delta^\gamma _\alpha \delta^\sigma _\nu \delta^\delta _\beta + \delta^\gamma _\alpha \delta^\delta _\nu \delta^\sigma _\beta + \delta^\delta _\alpha \delta^\gamma _\beta \delta^\sigma _\nu \\ & \ \ \  - \delta^\delta _\alpha \delta^\sigma _\beta \delta^\gamma _\nu - \delta^\sigma _\alpha \delta^\gamma _\beta \delta^\delta _\nu +  \delta^\sigma _\alpha \delta^\delta _\beta \delta^\gamma _\nu \ , \label{levicivita_1contracted}
        \end{split} \\\epsilon_{\mu\nu\alpha\beta}\epsilon^{\mu\nu\delta\sigma} &= - 2 \delta^\sigma  _\beta  \delta^\delta  _\alpha + 2 \delta^\delta  _\beta  \delta^\sigma _\alpha \ . \label{levicivita_2contracted}
    \end{align}
\end{subequations}
All other contractions with differently ordered indices can be acquired from the above identities using the tensor's asymmetry under the exchange of any two neighboring indices. Now we transform Eq.~(\ref{sigma_dot}) from the $\Sigma^{\mu\nu}$ to the $s^\mu$ representation by first looking at the left-hand side. We multiply the expression by $\frac{1}{2\mu c}\epsilon_{\alpha\beta\mu\nu}u^\beta$ and substitute $\Sigma^{\mu\nu}$ with the identity on the right in Eq.~(\ref{dual_transf}), thereby contracting both moment tensor indices
\begin{equation} \label{eq:sigma_to_s_LHS}
    \begin{aligned}
        \frac{1}{2\mu c}\epsilon_{\alpha\beta\mu\nu}\frac{\mathrm{d}\Sigma^{\mu\nu}}{\mathrm{d}\tau}u^\beta &= \frac{1}{2c^2}\epsilon_{\alpha\beta\mu\nu}\epsilon^{\mu\nu\delta\sigma}\frac{\mathrm{d}(s_\delta u_\sigma)}{\mathrm{d}\tau}u^\beta \\
        &= \frac{1}{c^2}(\delta^\delta  _\beta  \delta^\sigma _\alpha- \delta^\sigma  _\beta  \delta^\delta  _\alpha)\frac{\mathrm{d}(s_\delta u_\sigma)}{\mathrm{d}\tau}u^\beta \\
        &= \frac{1}{c^2}\frac{\mathrm{d}(s_\beta u_\alpha-s_\alpha u_\beta)}{\mathrm{d}\tau}u^\beta  \\
        &= \frac{1}{c^2}\bigg( \frac{\mathrm{d}s_\beta}{\mathrm{d}\tau}u_\alpha u^\beta + s_\beta \frac{\mathrm{d}u_\alpha}{\mathrm{d}\tau}u^\beta \\ & \ \ \ - \frac{\mathrm{d}s_\alpha}{\mathrm{d}\tau} u_\beta u^\beta - s_\alpha \frac{\mathrm{d}u_\beta}{\mathrm{d}\tau} u^\beta\bigg) \\
        &= - \frac{\mathrm{d}s_\alpha}{\mathrm{d}\tau} + \frac{1}{c^2}\frac{\mathrm{d}s_\beta}{\mathrm{d}\tau} u^\beta u_\alpha \\
        &= - \frac{\mathrm{d}s_\alpha}{\mathrm{d}\tau} - \frac{1}{c^2} s_\beta \frac{\mathrm{d}u^\beta}{\mathrm{d}\tau}  u_\alpha \\
        &= - \frac{\mathrm{d}s_\alpha}{\mathrm{d}\tau} + \frac{e}{mc^2} F _{\nu\beta}u^\nu s^\beta   u_\alpha \ .
    \end{aligned}
\end{equation}
The following steps were made in the equation above. In the second line we have used that $\epsilon_{\alpha\beta\mu\nu} = \epsilon_{\mu\nu\alpha\beta}$ and substituted Eq.~(\ref{levicivita_2contracted}), and have then applied the emerging Kronecker deltas to the term $(s_\delta u_\sigma)$ in the next one. In the fourth step, the product rule has been performed for the proper time derivative. In the following one, we have used the fact that $s_\beta u^\beta = 0$ and $\frac{\mathrm{d}u_\beta}{\mathrm{d}\tau} u^\beta = 0$, i.e., Eqs.~(\ref{su=0}) and~(\ref{eq:u_udot=0}), to remove the second and fourth term in the parenthesis as well as the norm of the four-velocity in the third term, $u_\beta u^\beta = c^2$. In the next to last line we have reformulated the second term by taking the proper time derivative of Eq.~(\ref{su=0}) and getting $\frac{\mathrm{d}s_\beta}{\mathrm{d}\tau} u^\beta = -s_\beta \frac{\mathrm{d}u^\beta}{\mathrm{d}\tau} $. In the last line we have substituted $\frac{\mathrm{d}u^\beta}{\mathrm{d}\tau}$ using the Lorentz force, Eq.~(\ref{lorentz_force_covariant}) and reformulating the term as $F^\beta _{\ \nu}u^\nu s_\beta=-F _{\nu\beta}u^\nu s^\beta $.
\begin{widetext}
We now perform an analogous set of operations to the first term on the right-hand side of Eq.~(\ref{sigma_dot})
\begin{equation} \label{eq:sigma_to_s_RHS}
    \begin{aligned}
        &\frac{eg}{2m} \frac{1}{2\mu c}\epsilon_{\alpha\beta\mu\nu}u^\beta \big( F^\mu  _{\ \alpha}\Sigma^{\alpha\nu}  - F^\nu  _{\ \alpha}\Sigma^{\alpha\mu} \big) =  \frac{eg}{4mc^2} \big(\epsilon_{\alpha\beta\mu\nu} \epsilon^{\rho\nu\delta\sigma} F^\mu  _{\ \rho}s_\delta u_\sigma u^\beta - \epsilon_{\alpha\beta\mu\nu} \epsilon^{\rho\mu\delta\sigma} F^\nu  _{\ \rho}s_\delta u_\sigma u^\beta \big) \\
        &=\frac{eg}{4mc^2} \big( ( -\delta^\rho _\alpha \delta^\delta _\beta  \delta^\sigma _\mu 
        + \delta^\rho _\alpha \delta^\sigma _\beta \delta^\delta _\mu  
        + \delta^\delta _\alpha \delta^\rho _\beta \delta^\sigma _\mu 
        - \delta^\delta _\alpha \delta^\sigma _\beta \delta^\rho _\mu 
        - \delta^\sigma _\alpha \delta^\rho _\beta \delta^\delta _\mu 
        +  \delta^\sigma _\alpha \delta^\delta _\beta \delta^\rho _\mu )F^\mu  _{\ \rho}s_\delta u_\sigma u^\beta \\
        & \ \ \ \ \ \ \ \ \ \ \ \ \, - (-\delta^\rho _\alpha \delta^\delta _\beta \delta^\sigma _\nu  
        + \delta^\rho _\alpha \delta^\sigma _\beta \delta^\delta _\nu  
        + \delta^\delta _\alpha \delta^\rho _\beta \delta^\sigma _\nu 
        - \delta^\delta _\alpha \delta^\sigma _\beta \delta^\rho _\nu 
        - \delta^\sigma _\alpha \delta^\rho _\beta \delta^\delta _\nu 
        +  \delta^\sigma _\alpha \delta^\delta _\beta \delta^\rho _\nu) F^\nu  _{\ \rho}s_\delta u_\sigma u^\beta \big)\\
        &= \frac{eg}{4mc^2} \big(  -F^\mu  _{\ \alpha}s_\beta u_\mu u^\beta  + F^\mu  _{\ \alpha}s_\mu u_\beta u^\beta + F^\mu  _{\ \beta}s_\alpha u_\mu u^\beta - F^\mu  _{\ \mu}s_\alpha u_\beta u^\beta - F^\mu  _{\ \beta}s_\mu u_\alpha u^\beta + F^\mu  _{\ \mu}s_\beta u_\alpha u^\beta \\
        &\ \ \ \ \ \ \ \ \ \ \ \ \ - F^\nu  _{\ \alpha}s_\beta u_\nu u^\beta  + F^\nu  _{\ \alpha}s_\nu u_\beta u^\beta + F^\nu  _{\ \beta}s_\alpha u_\nu u^\beta - F^\nu  _{\ \nu}s_\alpha u_\beta u^\beta - F^\nu  _{\ \beta}s_\nu u_\alpha u^\beta + F^\nu  _{\ \nu}s_\beta u_\alpha u^\beta \big) \\
        &= \frac{eg}{4mc^2} \big( c^2 F^\mu  _{\ \alpha}s_\mu   - F^\mu  _{\ \beta}s_\mu u_\alpha u^\beta
        + c^2 F^\nu  _{\ \alpha}s_\nu  - F^\nu  _{\ \beta}s_\nu u_\alpha u^\beta \big) \\
        &= -\frac{eg}{2m} F _{\alpha\nu}s^\nu + \frac{eg}{2mc^2} F_{\beta\nu}s^\nu u^\beta u _\alpha \ .
    \end{aligned}
\end{equation}
\end{widetext}
In the above manipulation, the expression is multiplied by $\frac{1}{2\mu c}\epsilon_{\alpha\beta\mu\nu}u^\beta$ as the left-hand side before and then the identity on the right in Eq.~(\ref{dual_transf}) is substituted. The Levi-Civita contractions from Eq.~(\ref{levicivita_1contracted}) are used in the second step after identifying $\epsilon_{\alpha\beta\mu\nu} \epsilon^{\rho\nu\delta\sigma} = \epsilon_{\nu\alpha\beta\mu} \epsilon^{\nu\rho\delta\sigma}$ and $\epsilon_{\alpha\beta\mu\nu} \epsilon^{\rho\mu\delta\sigma} = -\epsilon_{\mu\alpha\beta\nu} \epsilon^{\mu\rho\delta\sigma}$, and have been applied in the next one. In the next to last line, the following properties have been used to simplify the expression: $u_\mu u^\mu = c^2$, $s_\mu s^\mu = 0$, $F^\mu _{\ \mu} = 0$ and $F^\mu _{\ \beta} u_\mu u^\beta = 0$. Finally, to get to the last line in the above equation, we have used that $F^\mu  _{\ \alpha}s_\mu = F^\nu  _{\ \alpha}s_\nu = -F _{\alpha\nu}s^\nu$ and $F^\mu  _{\ \beta}s_\mu u^\beta = F^\nu  _{\ \beta}s_\nu u^\beta = -F_{\beta\nu}s^\nu u^\beta$.

Lastly, we perform the equivalent calculation of the second term on the right-hand side of Eq.~(\ref{sigma_dot})
\begin{widetext}
\begin{equation}
    \begin{aligned}
        &\frac{e(g-2)}{2mc^2} \frac{1}{2\mu c}\epsilon_{\alpha\beta\mu\nu}u^\beta \big(u^\mu\Sigma^{\nu\alpha} - u^\nu \Sigma^{\mu\alpha}\big)F_{\alpha\rho}u^\rho \\ &= \frac{e(g-2)}{4mc^4} \big( \epsilon_{\alpha\beta\mu\nu} \epsilon^{\nu\alpha\delta\sigma} F_{\alpha\rho} s_\delta u_\sigma  u^\beta u ^\mu u^\rho - \epsilon_{\alpha\beta\mu\nu} \epsilon^{\mu\alpha\delta\sigma} F_{\alpha\rho} s_\delta u_\sigma  u^\beta u ^\nu u^\rho \big) \\
        &= \frac{e(g-2)}{4mc^4} \big( ( 2\delta ^\sigma _\mu \delta^\delta _\beta - 2\delta ^\delta _\mu \delta ^\sigma _\beta) F_{\alpha\rho} s_\delta u_\sigma  u^\beta u ^\mu u^\rho - ( - 2\delta ^\sigma _\nu \delta^\delta _\beta + 2\delta ^\delta _\nu \delta ^\sigma _\beta) F_{\alpha\rho} s_\delta u_\sigma  u^\beta u ^\nu u^\rho \big) \\
        &= \frac{e(g-2)}{2mc^4} \big(  F_{\alpha\rho} s_\beta u_\mu  u^\beta u ^\mu u^\rho - F_{\alpha\rho} s_\mu u_\beta  u^\beta u ^\mu u^\rho + F_{\alpha\rho} s_\beta u_\nu  u^\beta u ^\nu u^\rho - F_{\alpha\rho} s_\nu u_\beta  u^\beta u ^\nu u^\rho \big) \\ &=0 \ .        
    \end{aligned}
\end{equation}    
\end{widetext}
In the second line above the identity for $\Sigma^{\mu\nu}$ from Eq.~(\ref{dual_transf}) has been substituted and the Levi-Civita contractions in Eq.~(\ref{levicivita_2contracted}) utilized after reformulating $\epsilon_{\alpha\beta\mu\nu} \epsilon^{\nu\alpha\delta\sigma} = - \epsilon_{\nu\alpha\beta\mu} \epsilon^{\nu\alpha\delta\sigma}$ and $\epsilon_{\alpha\beta\mu\nu} \epsilon^{\mu\alpha\delta\sigma} = \epsilon_{\mu\alpha\beta\nu} \epsilon^{\mu\alpha\delta\sigma}$. In the fourth line, the Kronecker deltas have been applied, where we see that in every term there is a contraction between the spin-pseudovector $s^\mu$ and the four-velocity $u^\mu$, which means that each term and, in turn, the whole expression vanishes and does not contribute to the $s^\mu$ representation of the evolution equation for the magnetic sources. Putting together Eqs.~(\ref{eq:sigma_to_s_LHS}) and~(\ref{eq:sigma_to_s_RHS}) and rearranging them to isolate $\frac{\mathrm{d}s_\alpha}{\mathrm{d}\tau}$ on the left-hand side yields
\begin{equation}
    \begin{aligned}
        \frac{\mathrm{d}s_\alpha}{\mathrm{d}\tau} &= \frac{eg}{2m} F _{\alpha\nu}s^\nu - \frac{eg}{2mc^2} F_{\beta\nu}s^\nu u^\beta u _\alpha + \frac{e}{mc^2} F _{\nu\beta}u^\nu s^\beta   u_\alpha \\
        &= \frac{eg}{2m} F _{\alpha\nu}s^\nu - \frac{e(g-2)}{2mc^2} u _\alpha F_{\beta\nu} u^\beta s^\nu \ ,
    \end{aligned}
\end{equation}
the result of which corresponds to Eq.~(\ref{eq:sdot}) for the contravariant vector $s_\mu$. To get to the final expression, the dummy indices in the last term on the right-hand side of the first line have been exchanged, which makes obvious the fact that the last two terms are identical up to a prefactor, which, when put together, gives the desired $(g-2)$-term in the result. The conversion of the equation of motion for the magnetic sources from the $s^\mu$ to the $\Sigma^{\mu\nu}$ representation works in an equivalent way.

\bibliography{main}

\end{document}